\documentclass[amssymb,amsmath,prd,twocolumn,superscriptaddress,nofootinbib]{revtex4-1}

\usepackage{graphicx}
\usepackage{bm}
\usepackage{amssymb,amsmath}
\usepackage{mathrsfs}
\usepackage{latexsym}
\usepackage{color}
\usepackage[normalem]{ulem} 
\usepackage{dcolumn}
\usepackage[colorlinks=true,citecolor=blue,urlcolor=blue]{hyperref}
\usepackage[usenames,dvipsnames]{xcolor}
\usepackage{multirow}
\usepackage{comment}
\usepackage{lineno}

\newcommand{\beq}{\begin{equation}}
\newcommand{\eeq}{\end{equation}}
\newcommand{\ba}{\begin{array}}
\newcommand{\ea}{\end{array}}
\newcommand{\bea}{\begin{eqnarray}}
\newcommand{\eea}{\end{eqnarray}}
\newcommand{\bc}{\begin{center}}
\newcommand{\ec}{\end{center}}

\newcommand{\aap}{Astron.~Astrophys.}
\newcommand{\nphysa}{Nuc.~Phys.~A}

\newcommand{\Msolar}{{\rm M}_{\odot}}
\newcommand{\Mmax}{M_{\rm max}}

\newcommand{\nn}{\nonumber}

\allowdisplaybreaks

\definecolor{green}{rgb}{0.03,0.47,0.19}
\definecolor{magenta}{rgb}{0.54, 0.17, 0.88}
\definecolor{orange}{rgb}{0.851, 0.372, 0.007}

\newcommand{\dgw}{\ensuremath{d_\mathrm{GW}}}
\newcommand{\dnicer}{\ensuremath{d_\mathrm{X}}}
\newcommand{\dmoi}{\ensuremath{d_\mathrm{I}}}
\newcommand{\dm}{\ensuremath{d_\mathrm{M}}}

\newcommand{\Lgw}{\ensuremath{\lambda_\mathrm{GW}}}
\newcommand{\Lnicer}{\ensuremath{\lambda_\mathrm{X}}}
\newcommand{\Lmoi}{\ensuremath{\lambda_\mathrm{I}}}
\newcommand{\Lm}{\ensuremath{\lambda_\mathrm{M}}}

\newcommand{\CCA}{\affiliation{Center for Computational Astrophysics, Flatiron Institute, 162 5th Avenue, New York, New York 10010, USA}}

\newcommand{\result}[1]{\textcolor{black}{#1}}

\newcommand{\PriorMmax}{\result{\ensuremath{1.48^{+0.73}_{-1.38}}}} 
\newcommand{\PriorR}{\result{\ensuremath{8.12^{+5.81}_{-3.98}}}}    
\newcommand{\PriorL}{\result{\ensuremath{25^{+899}_{-25}}}}    
\newcommand{\PriorI}{\result{\ensuremath{0.79^{+1.05}_{-0.31}}}}    

\newcommand{\PSRsMmax}{\result{\ensuremath{2.27^{+0.46}_{-0.27}}}}  
\newcommand{\PSRsR}{\result{\ensuremath{13.64^{+2.97}_{-3.23}}}}     
\newcommand{\PSRsL}{\result{\ensuremath{823^{+1450}_{-823}}}}     
\newcommand{\PSRsI}{\result{\ensuremath{1.78^{+0.66}_{-0.66}}}}     

\newcommand{\PSRsGWsMmax}{\result{\ensuremath{2.20^{+0.24}_{-0.18}}}} 
\newcommand{\PSRsGWsR}{\result{\ensuremath{10.95^{+2.00}_{-1.37}}}} 
\newcommand{\PSRsGWsL}{\result{\ensuremath{228^{+319}_{-134}}}} 
\newcommand{\PSRsGWsI}{\result{\ensuremath{1.28^{+0.35}_{-0.19}}}} 

\newcommand{\PSRsXrayMmax}{\result{\ensuremath{2.25^{+0.38}_{-0.25}}}} 
\newcommand{\PSRsXrayR}{\result{\ensuremath{13.38^{+1.40}_{-1.69}}}} 
\newcommand{\PSRsXrayL}{\result{\ensuremath{749^{+550}_{-500}}}} 
\newcommand{\PSRsXrayI}{\result{\ensuremath{1.73^{+0.33}_{-0.34}}}} 

\newcommand{\AllMmax}{\result{\ensuremath{2.22^{+0.30}_{-0.20}}}}  
\newcommand{\AllR}{\result{\ensuremath{12.32^{+1.09}_{-1.47}}}}     
\newcommand{\AllL}{\result{\ensuremath{451^{+241}_{-279}}}}     
\newcommand{\AllI}{\result{\ensuremath{1.51^{+0.20}_{-0.30}}}}     

\newcommand{\CurrentDR}{\result{\ensuremath{2.56}}} 

\newcommand{\PnucCoef}{\ensuremath{10^{33}}}
\newcommand{\PtwonucCoef}{\ensuremath{10^{34}}}

\newcommand{\PsixnucCoef}{\ensuremath{10^{35}}}

\newcommand{\PriorPnuc}{\result{\ensuremath{2.3^{+6.5}_{-2.2}}}} 
\newcommand{\PriorPtwonuc}{\result{\ensuremath{1.2^{+5.0}_{-1.2}}}} 
\newcommand{\PriorPsixnuc}{\result{\ensuremath{2.5^{+4.8}_{-2.5}}}} 

\newcommand{\PSRsPnuc}{\result{\ensuremath{6.4^{+9.3}_{-6.3}}}} 
\newcommand{\PSRsPtwonuc}{\result{\ensuremath{6.2^{+4.8}_{-6.2}}}} 
\newcommand{\PSRsPsixnuc}{\result{\ensuremath{7.6^{+6.9}_{-5.5}}}} 

\newcommand{\PSRsGWsPnuc}{\result{\ensuremath{2.2^{+4.4}_{-2.1}}}} 
\newcommand{\PSRsGWsPtwonuc}{\result{\ensuremath{1.8^{+3.0}_{-1.8}}}} 
\newcommand{\PSRsGWsPsixnuc}{\result{\ensuremath{9.1^{+4.1}_{-4.0}}}} 

\newcommand{\PSRsXrayPnuc}{\result{\ensuremath{5.7^{+4.1}_{-5.2}}}} 
\newcommand{\PSRsXrayPtwonuc}{\result{\ensuremath{6.0^{+4.4}_{-3.9}}}} 
\newcommand{\PSRsXrayPsixnuc}{\result{\ensuremath{7.4^{+6.6}_{-4.6}}}} 

\newcommand{\AllPnuc}{\result{\ensuremath{4.3^{+3.8}_{-4.0}}}} 
\newcommand{\AllPtwonuc}{\result{\ensuremath{3.8^{+2.7}_{-2.9}}}} 
\newcommand{\AllPsixnuc}{\result{\ensuremath{8.6^{+5.3}_{-4.3}}}} 

\newcommand{\PriorMaxCs}{\result{\ensuremath{0.76^{+0.25}_{-0.38}}}} 
\newcommand{\PriorRhoMaxCs}{\result{\ensuremath{1.31^{+1.59}_{-1.28}}}} 
\newcommand{\PriorPMaxCs}{\result{\ensuremath{1.5^{+8.2}_{-1.5}}}} 

\newcommand{\PSRsMaxCs}{\result{\ensuremath{0.74^{+0.26}_{-0.28}}}} 
\newcommand{\PSRsRhoMaxCs}{\result{\ensuremath{0.93^{+0.65}_{-0.70}}}} 
\newcommand{\PSRsPMaxCs}{\result{\ensuremath{2.5^{+5.3}_{-2.5}}}} 

\newcommand{\PSRsGWsMaxCs}{\result{\ensuremath{0.93^{+0.07}_{-0.23}}}} 
\newcommand{\PSRsGWsRhoMaxCs}{\result{\ensuremath{1.20^{+0.76}_{-0.59}}}} 
\newcommand{\PSRsGWsPMaxCs}{\result{\ensuremath{4.1^{+8.6}_{-4.0}}}} 

\newcommand{\PSRsXrayMaxCs}{\result{\ensuremath{0.70^{+0.30}_{-0.23}}}} 
\newcommand{\PSRsXrayRhoMaxCs}{\result{\ensuremath{1.00^{+0.58}_{-0.63}}}} 
\newcommand{\PSRsXrayPMaxCs}{\result{\ensuremath{2.9^{+4.7}_{-2.8}}}} 

\newcommand{\AllMaxCs}{\result{\ensuremath{0.85^{+0.15}_{-0.29}}}} 
\newcommand{\AllRhoMaxCs}{\result{\ensuremath{1.12^{+0.63}_{-0.64}}}} 
\newcommand{\AllPMaxCs}{\result{\ensuremath{3.6^{+6.8}_{-3.6}}}} 

\newcommand{\AllRonebranch}{\result{\ensuremath{12.36^{+1.44}_{-1.11}}}}
\newcommand{\AllRtwobranch}{\result{\ensuremath{11.86^{+1.08}_{-1.69}}}}

\newcommand{\BayesTwoOneBranch}{\result{\ensuremath{0.24\pm0.02}}} 
\newcommand{\BayesTwoOneBranchGivenPSR}{\result{\ensuremath{1.8\pm0.2}}} 

\newcommand{\OhEightMone}{\result{\ensuremath{1.48^{+0.13}_{-0.11}}}} 
\newcommand{\OhEightRone}{\result{\ensuremath{11.04^{+2.06}_{-1.33}}}} 
\newcommand{\OhEightLone}{\result{\ensuremath{178^{+318}_{-144}}}} 
\newcommand{\UpdatedOhEightMone}{\result{\ensuremath{1.46^{+0.13}_{-0.09}}}} 
\newcommand{\UpdatedOhEightRone}{\result{\ensuremath{12.33^{+1.10}_{-1.45}}}} 
\newcommand{\UpdatedOhEightLone}{\result{\ensuremath{332^{+295}_{-249}}}} 

\newcommand{\OhEightMtwo}{\result{\ensuremath{1.27^{+0.10}_{-0.10}}}} 
\newcommand{\OhEightRtwo}{\result{\ensuremath{10.94^{+2.23}_{-1.51}}}} 
\newcommand{\OhEightLtwo}{\result{\ensuremath{447^{+593}_{-299}}}} 
\newcommand{\UpdatedOhEightMtwo}{\result{\ensuremath{1.28^{+0.08}_{-0.10}}}} 
\newcommand{\UpdatedOhEightRtwo}{\result{\ensuremath{12.29^{+1.20}_{-1.49}}}} 
\newcommand{\UpdatedOhEightLtwo}{\result{\ensuremath{737^{+544}_{-474}}}} 

\newcommand{\OhFourMone}{\result{\ensuremath{2.02^{+0.49}_{-0.37}}}} 
\newcommand{\OhFourRone}{\result{\ensuremath{12.18^{+2.31}_{-5.83}}}} 
\newcommand{\OhFourLone}{\result{\ensuremath{43^{+178}_{-43}}}} 
\newcommand{\UpdatedOhFourMone}{\result{\ensuremath{2.03^{+0.46}_{-0.38}}}} 
\newcommand{\UpdatedOhFourRone}{\result{\ensuremath{11.79^{+1.49}_{-5.21}}}} 
\newcommand{\UpdatedOhFourLone}{\result{\ensuremath{31^{+108}_{-31}}}} 

\newcommand{\OhFourMtwo}{\result{\ensuremath{1.36^{+0.26}_{-0.25}}}} 
\newcommand{\OhFourRtwo}{\result{\ensuremath{12.90^{+2.18}_{-2.97}}}} 
\newcommand{\OhFourLtwo}{\result{\ensuremath{648^{+1770}_{-648}}}} 
\newcommand{\UpdatedOhFourMtwo}{\result{\ensuremath{1.35^{+0.28}_{-0.23}}}} 
\newcommand{\UpdatedOhFourRtwo}{\result{\ensuremath{12.31^{+1.11}_{-1.51}}}} 
\newcommand{\UpdatedOhFourLtwo}{\result{\ensuremath{521^{+912}_{-461}}}} 

\newcommand{\NicerM}{\result{\ensuremath{1.48^{+0.20}_{-0.23}}}} 
\newcommand{\NicerR}{\result{\ensuremath{13.38^{+1.47}_{-1.67}}}} 
\newcommand{\NicerL}{\result{\ensuremath{539^{+378}_{-296}}}} 
\newcommand{\UpdatedNicerM}{\result{\ensuremath{1.36^{+0.17}_{-0.20}}}} 
\newcommand{\UpdatedNicerR}{\result{\ensuremath{12.28^{+1.09}_{-1.53}}}} 
\newcommand{\UpdatedNicerL}{\result{\ensuremath{497^{+369}_{-274}}}} 

\newcommand{\AltPSRsXrayMmax}{\result{\ensuremath{2.25^{+0.37}_{-0.24}}}} 
\newcommand{\AltPSRsXrayR}{\result{\ensuremath{13.21^{+1.33}_{-1.69}}}} 
\newcommand{\AltPSRsXrayL}{\result{\ensuremath{698^{+458}_{-479}}}} 
\newcommand{\AltPSRsXrayI}{\result{\ensuremath{1.70^{+0.30}_{-0.35}}}} 

\newcommand{\AltAllMmax}{\result{\ensuremath{2.22^{+0.28}_{-0.21}}}}  
\newcommand{\AltAllR}{\result{\ensuremath{12.06^{+1.26}_{-1.43}}}}     
\newcommand{\AltAllL}{\result{\ensuremath{400^{+259}_{-253}}}}     
\newcommand{\AltAllI}{\result{\ensuremath{1.47^{+0.24}_{-0.27}}}}     

\newcommand{\AltPSRsXrayPnuc}{\result{\ensuremath{5.3^{+3.6}_{-5.0}}}} 
\newcommand{\AltPSRsXrayPtwonuc}{\result{\ensuremath{5.7^{+4.2}_{-4.1}}}} 
\newcommand{\AltPSRsXrayPsixnuc}{\result{\ensuremath{7.5^{+6.5}_{-4.6}}}} 

\newcommand{\AltAllPnuc}{\result{\ensuremath{4.0^{+3.8}_{-3.8}}}} 
\newcommand{\AltAllPtwonuc}{\result{\ensuremath{3.4^{+2.7}_{-2.8}}}} 
\newcommand{\AltAllPsixnuc}{\result{\ensuremath{8.8^{+4.9}_{-4.4}}}} 

\newcommand{\AltPSRsXrayMaxCs}{\result{\ensuremath{0.72^{+0.28}_{-0.25}}}} 
\newcommand{\AltPSRsXrayRhoMaxCs}{\result{\ensuremath{1.02^{+0.57}_{-0.64}}}} 
\newcommand{\AltPSRsXrayPMaxCs}{\result{\ensuremath{3.0^{+5.1}_{-2.9}}}} 

\newcommand{\AltAllMaxCs}{\result{\ensuremath{0.87^{+0.13}_{-0.29}}}} 
\newcommand{\AltAllRhoMaxCs}{\result{\ensuremath{1.14^{+0.66}_{-0.64}}}} 
\newcommand{\AltAllPMaxCs}{\result{\ensuremath{3.7^{+7.4}_{-3.7}}}} 

\newcommand{\MockCurrentDR}{\result{\ensuremath{3.20}}} 

\newcommand{\MockOFourDR}{\result{\ensuremath{2.12}}} 
\newcommand{\MockOFourDRFraction}{\result{\ensuremath{34\%}}} 

\newcommand{\MockOFourNICERDR}{\result{\ensuremath{1.72}}} 
\newcommand{\MockOFourNICERFraction}{\result{\ensuremath{46\%}}} 

\newcommand{\MockDesignDR}{\result{\ensuremath{1.17}}} 

\newcommand{\MockDesignNICERDR}{\result{\ensuremath{1.07}}} 

\newcommand{\MockCurrentMOIDR}{\result{\ensuremath{2.33}}} 
\newcommand{\MockCurrentMOIFraction}{\result{\ensuremath{27\%}}} 

\newcommand{\MockDesignNICERMOIPtwonucRelError}{\result{\ensuremath{80\%}}} 

\begin{document}

\title{
Nonparametric constraints on neutron star matter with existing and upcoming gravitational wave and pulsar observations
}

\author{Philippe Landry}\email{plandry@fullerton.edu}
\affiliation{Gravitational-Wave Physics \& Astronomy Center, California State University, Fullerton, 800 N State College Boulevard, Fullerton, California 92831, USA}

\author{Reed Essick}\email{reed.essick@gmail.com}
\affiliation{Kavli Institute for Cosmological Physics, University of Chicago, 5640 S Ellis Avenue, Chicago, Illinois 60637, USA}

\author{Katerina Chatziioannou}\email{kchatziioannou@flatironinstitute.org}
\CCA

\date{\today}

\begin{abstract}
Observations of neutron stars, whether in binaries or in isolation, provide information about the internal structure of the most extreme material objects in the Universe.
In this work, we combine information from recent observations to place joint constraints on the properties of neutron star matter.
We use (i) lower limits on the maximum mass of neutron stars obtained through radio observations of heavy pulsars, (ii) constraints on tidal properties inferred through the gravitational waves neutron star binaries emit as they coalesce, and (iii) information about neutron stars' masses and radii obtained through X-ray emission from surface hot spots.
In order to combine information from such distinct messengers while avoiding the kind of modeling systematics intrinsic to parametric inference schemes, we employ a nonparametric representation of the neutron-star equation of state based on Gaussian processes conditioned on nuclear theory models.
We find that existing astronomical observations imply $R_{1.4}=\AllR\,$km for the radius of a $1.4\,\Msolar$ neutron star and $p(2\rho_\mathrm{nuc})=\AllPtwonuc\times\PtwonucCoef\,\mathrm{dyn}/\mathrm{cm}^2$ for the pressure at twice nuclear saturation density at the 90\% credible level.
The upper bounds are driven by the gravitational wave observations, while X-ray and heavy pulsar observations drive the lower bounds.
Additionally, we compute expected constraints from potential future astronomical observations and find that they can jointly determine $R_{1.4}$ to ${\cal{O}}(1)\,$km and $p(2\rho_\mathrm{nuc})$ to \MockDesignNICERMOIPtwonucRelError~relative uncertainty in the next five years.
\end{abstract}

\maketitle
  
\section{Introduction}
\label{sec:intro}

Despite decades of theoretical, experimental, and observational work, the properties and composition of the cold, extremely dense matter inside neutron stars (NSs) are still uncertain~\cite{Lattimer:2015nhk,Oertel:2016bki,Baym:2017whm}.
Terrestrial experiments, such as heavy-ion collisions~\cite{Danielewicz:2002pu}, typically probe lower densities than those encountered in NS cores.
Observations of X-ray binaries involving accreting NSs have been used to constrain their radii, though astrophysical systematics may impact the estimates~\cite{Ozel:2016oaf,Miller:2016pom}.
At the same time, surveys of massive pulsars place lower limits on the maximum mass a nonrotating NS can attain, and suggest that the equation of state (EoS) of NSs must be stiff enough to support $\sim2\,\Msolar$ stars with radii of ${\cal{O}}(10)\,$km~\cite{Demorest:2010bx,Antoniadis:2013pzd,Cromartie:2019kug}.

The past few years have seen the emergence of novel observational methods for constraining the properties of NSs.
Gravitational wave (GW) observations of coalescing NSs with LIGO~\cite{TheLIGOScientific:2014jea} and Virgo~\cite{TheVirgo:2014hva} offer the possibility of measuring the tidal properties of inspiraling binary NSs (BNSs), quantified through the tidal deformability. 
Depending on the distribution of NS masses in coalescing systems throughout the Universe, GWs may potentially be used to probe a wide range of masses and therefore central densities~\cite{Read:2009yp,Hinderer:2009ca,Lackey:2014fwa,DelPozzo:2013ala,Chatziioannou:2015uea}. 
Additionally, X-ray measurements of emission from hot spots on the NS surface with NICER~\cite{Watts:2016uzu} can offer information about the mass and radius of selected pulsars, either in isolation or in binaries.
The characteristics and shape of the observed pulses carry an imprint of the NS spacetime in which the hot spot emission propagates, the properties of which are determined primarily by the compactness of the star, i.e. the dimensionless ratio of its mass to its radius~\cite{Miller:2014mca,Watts:2016uzu}.
Other possible future probes of NS structure include moment of inertia (MoI) measurements via pulsar timing~\cite{Lattimer_2005,Landry:2018jyg}, postmerger signals from BNS coalescences~\cite{bauswein:15,Baiotti:2016qnr,Chatziioannou:2017ixj,Torres-Rivas:2018svp}, coalescences of NSs and black holes (BHs)~\cite{PhysRevD.85.044061,Lackey:2013axa,Kumar:2016zlj}, or pulsar spin measurements~\cite{Hessels2006}.

To date, two GW signals likely originating from the coalescence of BNSs have been announced, GW170817~\cite{TheLIGOScientific:2017qsa} and GW190425~\cite{Abbott:2020uma}. 
The former constitutes the first detected GW signal for which matter effects are present, and the first with a luminous counterpart observed across the electromagnetic spectrum~\cite{GBM:2017lvd}. 
With a chirp mass of $\approx 1.2\,\Msolar$ and a signal-to-noise ratio (SNR) of 32.4~\cite{Abbott:2018wiz}, GW170817 is consistent with the population of Galactic double NS binaries and has already been extensively used to study the properties of dense matter, e.g.~\cite{Margalit:2017dij,Annala:2017llu,Bauswein:2017vtn,Shibata:2017xdx,Radice:2017lry,Ruiz:2017due,Raithel:2018ncd,De:2018uhw,Abbott:2018exr,Most:2018hfd,Radice:2018ozg,Coughlin:2018miv,2018ApJ...852L..25R,Lim:2018bkq,Ayriyan:2018blj,Coughlin:2018fis,Landry:2018prl,Essick:2019ldf,LimHolt2019,JiangTang2019,CapanoTews2019,KiuchiKyutoku2019,Shibata:2019ctb,ShaoTang2020}. GW190425 originates from a more massive system, with a chirp mass of $\approx 1.44\,\Msolar$, and is quieter, with an SNR of 12.9~\cite{Abbott:2020uma}.
The discovery of GW190425 is seemingly promising for GW studies of dense matter, as more massive (and therefore more compact) stars can be used to probe higher densities.
However, the weaker tidal interactions experienced by more massive stars, coupled with the low SNR of the system, result in the signal being less informative than GW170817~\cite{Abbott:2020uma} from the dense-matter point of view.

NICER commenced observations in 2017 and recently announced mass and radius constraints for its first target, PSR J0030+0451~\cite{Guillot:2019vqp,Bogdanov:2019ixe,Bogdanov:2019qjb}.
Independent analyses found X-ray pulses consistent with emission from two noncircular hot spots in the southern hemisphere of the pulsar~\cite{Miller:2019cac,Riley:2019yda}, with weak support for a third oval spot~\cite{Miller:2019cac}. The shape of the pulses was modeled with an oblate Schwarzschild model for the NS~\cite{MorsinkLeahy2007,NattilaPihajoki2018}, which includes a number of propagation effects such as Doppler redshift, gravitational redshift, and aberration.
The resulting constraints on the mass and radius of J0030+0451 point to a star with a radius of $\approx 13\,$km with an uncertainty of about $4\,$km at the $90\%$ credible level, and a mass of $\approx 1.4\,\Msolar$.\footnote{When quoting absolute and relative uncertainties, we report the full width of the 90\% credible interval unless otherwise specified.} These measurements, in turn, inform constraints on the dense-matter EoS~\cite{Miller:2019cac,Raaijmakers:2019qny}.

In anticipation of these observations, a number of techniques to combine information from multiple sources have been suggested in order to achieve overall stronger constraints on NS properties \cite{DelPozzo:2013ala,Lackey:2014fwa,Miller:2019nzo,Forbes:2019xaz}.
These approaches make use of the fact that all NSs in the Universe are expected to share a common EoS, i.e. a common relation between their internal pressure and density.
Coupled to the structure equations for NSs~\cite{Tolman1939,OppenheimerVolkoff1939,Hartle1967,Hinderer:2007mb}, the EoS uniquely determines the relation between the stellar radius and mass, as well as the tidal deformability and other macroscopic observables. 
However, combining observations in this way requires a generic means of representing the uncertain density as a function of the pressure (or vice versa).

Several different parametrizations of the EoS have been proposed for this purpose and have subsequently been used to infer the EoS from astronomical observations.
In the case of EoSs without strong phase transitions, the tidal deformability as a function of mass can be expressed through a Taylor expansion about a fiducial NS mass~\cite{DelPozzo:2013ala}. 
More generic parametrizations in terms of piecewise polytropes~\cite{Read:2008iy,Raithel:2016bux} and a spectral decomposition~\cite{Lindblom:2010bb,2012PhRvD..86h4003L,Lindblom:2013kra,Lindblom:2018rfr} have also been extensively used~\cite{Lackey:2014fwa,Raithel:2017ity,Carney:2018sdv,Abbott:2018exr,Vivanco:2019qnt} due to their generality and flexibility.
Likewise, parametrizations inspired by nuclear physics frameworks have been explored, for example by combining low-density chiral effective field theory computations with generic high-density parametrizations~\cite{CapanoTews2019,Forbes:2019xaz, Dietrich:2020lps}.
Additionally, custom EoS parametrizations targeting phase transitions from hadronic to quark matter have been proposed~\cite{Alford:2013aca,Alford:2015gna,Greif:2018njt,Tews:2019cap} and studied in the context of current and future observations~\cite{Han:2018mtj,Montana:2018bkb,Chatziioannou:2019yko}.  

While many of these parametrizations have been used to study the impact of GW170817 or J0030+0451 individually, it is only very recently that they have been deployed to jointly analyze data from LIGO, Virgo and NICER. In~\cite{Miller:2019cac}, both piecewise-polytrope and spectral representations were used to constrain the EoS with combined information from J0030+0451's mass and radius (assuming a 3 hot spot configuration), GW170817's tidal deformabilities, and the masses of the three heaviest pulsars. 
Reference~\cite{Raaijmakers:2019dks} performed a similar analysis, but assumed a crescent-plus-oval hot spot geometry (ST+PST from~\cite{Riley:2019yda}) for J0030+0451, and used a constant sound-speed parametrization in place of the spectral one. Like Ref.~\cite{Miller:2019cac}, Ref.~\cite{Jiang:2019rcw} adopted piecewise-polytrope and spectral parametrizations, but approximated the mass-radius likelihood from~\cite{Riley:2019yda} as a two-dimensional Gaussian and modeled the heavy pulsar information as a step-function likelihood $\Theta(M_{\rm max} \geq 2.04\,\Msolar)$. Unlike other studies, they also folded constraints from an interpretation \cite{TewsLattimer2017} of terrestrial experiments into their prior. Ultimately, all three studies obtained broadly consistent constraints on the EoS, although only~\cite{Jiang:2019rcw} presented quantitative results detailed enough to permit close comparison.

In this study, we examine the combined effect of the latest astronomical measurements, including GW190425, on the inference of the NS EoS.
In contrast to previous joint analyses of LIGO, Virgo, and NICER data, we adopt the nonparametric representation developed in~\cite{Landry:2018prl,Essick:2019ldf} rather than a parametrization for the EoS.
The nonparametric approach allows for more model freedom in the EoS representation than any direct parametrization with a small number of parameters.
Using Gaussian processes conditioned on candidate EoSs from nuclear theory, we create a generative model that emulates all types of behavior within the class of causal and thermodynamically stable EoSs, including features such as strong phase transitions, without the need for explicit parametrizations.
The Gaussian process is tunable in that the generative model can closely follow existing theoretical proposals or fully explore the complete function space of physically allowable EoSs.
In this work, we use a \emph{model-agnostic} prior that tracks the candidate EoSs very loosely and explores the full set of causal and thermodynamically stable EoS.
The prior process is composed of three subprocesses conditioned on candidate EoSs of different compositions, namely hadronic ($npe\mu$), hyperonic ($npe\mu{Y}$), or quark ($npe\mu(Y)Q$), some of which support strong phase transitions and/or hybrid stars.
However, because the \emph{agnostic} prior is only loosely tied to the input EoSs, the distinctions between the subprocesses are small in practice.

Using our nonparametric EoS representation, we calculate constraints on the NS EoS based on three classes of astronomical observations: 
\begin{enumerate}
    \item Radio measurements of pulsar masses, in particular the heaviest known pulsars. We use the masses for PSR J1614$-$2230~\cite{Demorest:2010bx,Fonseca:2016tux}, PSR J0348+0432~\cite{Antoniadis:2013pzd}, and PSR J0740+6620~\cite{Cromartie:2019kug}.
    \item Measurements of the tidal properties of NSs probed through GWs emitted during the inspiral stage of NS coalescences. We use data from GW170817~\cite{TheLIGOScientific:2017qsa} and GW190425~\cite{Abbott:2020uma}.
    \item Measurements of pulsar masses and radii, obtained with X-ray timing of emission from surface hot spots with NICER. We use the recently published results for J0030+0451~\cite{Miller:2019cac,Riley:2019yda}.
\end{enumerate}
We refer the reader to Tables~\ref{tab:Maxmassdata},~\ref{tab:BNSdata}, and~\ref{tab:NICERdata} for further details.
Measurements of other macroscopic observables, such as NS spins, or constraints on microscopic physics, from either nuclear experiments or theoretical considerations, could also be incorporated into our inference scheme.
However, GWs, X-ray timing, and massive pulsars are currently the most relevant observations and are likely to continue to dominate in the near future.

\begin{figure*}[]
    \parbox{0.48\hsize}{
    \includegraphics[width=\hsize]{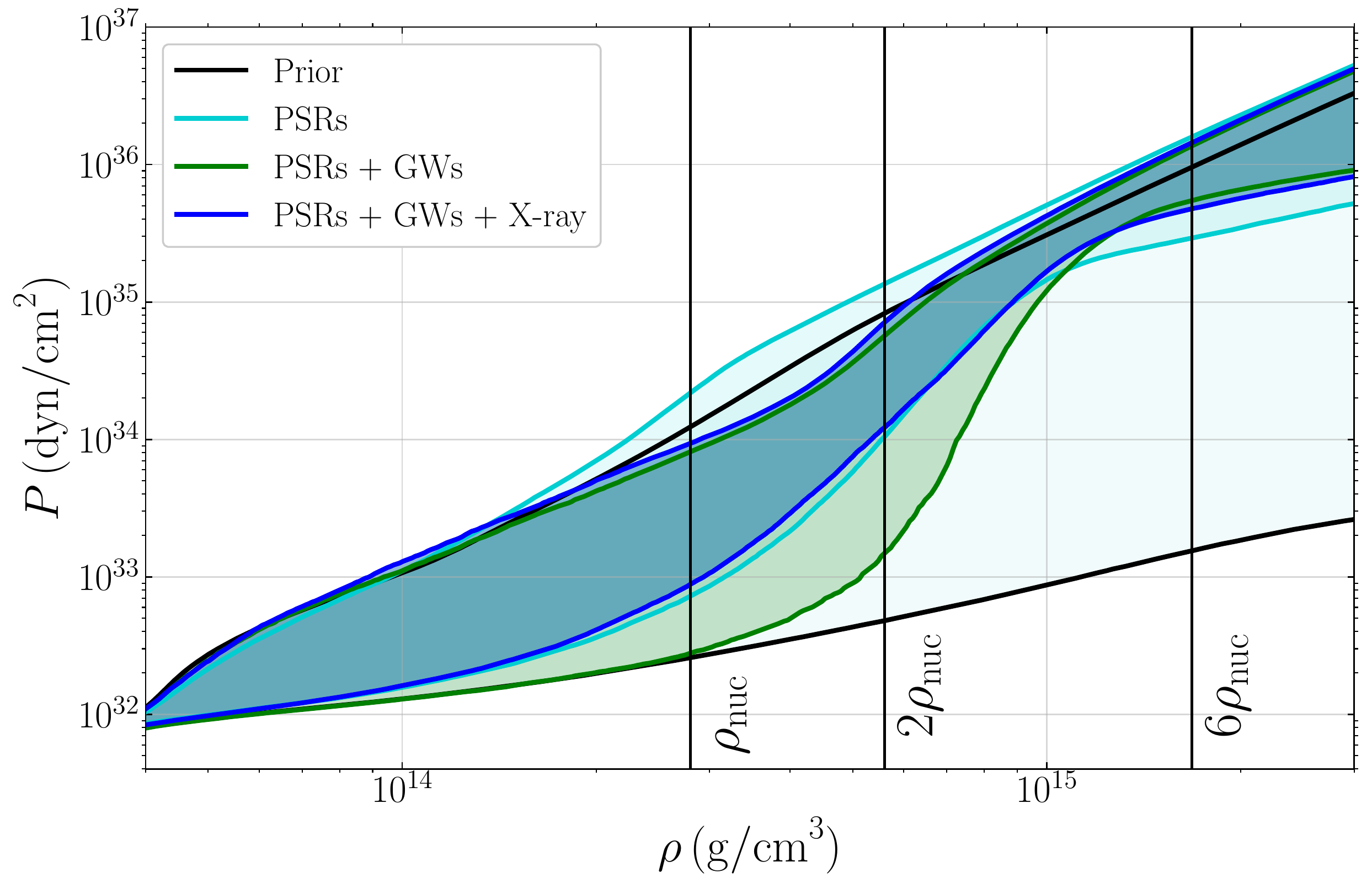}\\[-2ex]
    }\parbox{0.48\hsize}{
    \includegraphics[width=\hsize]{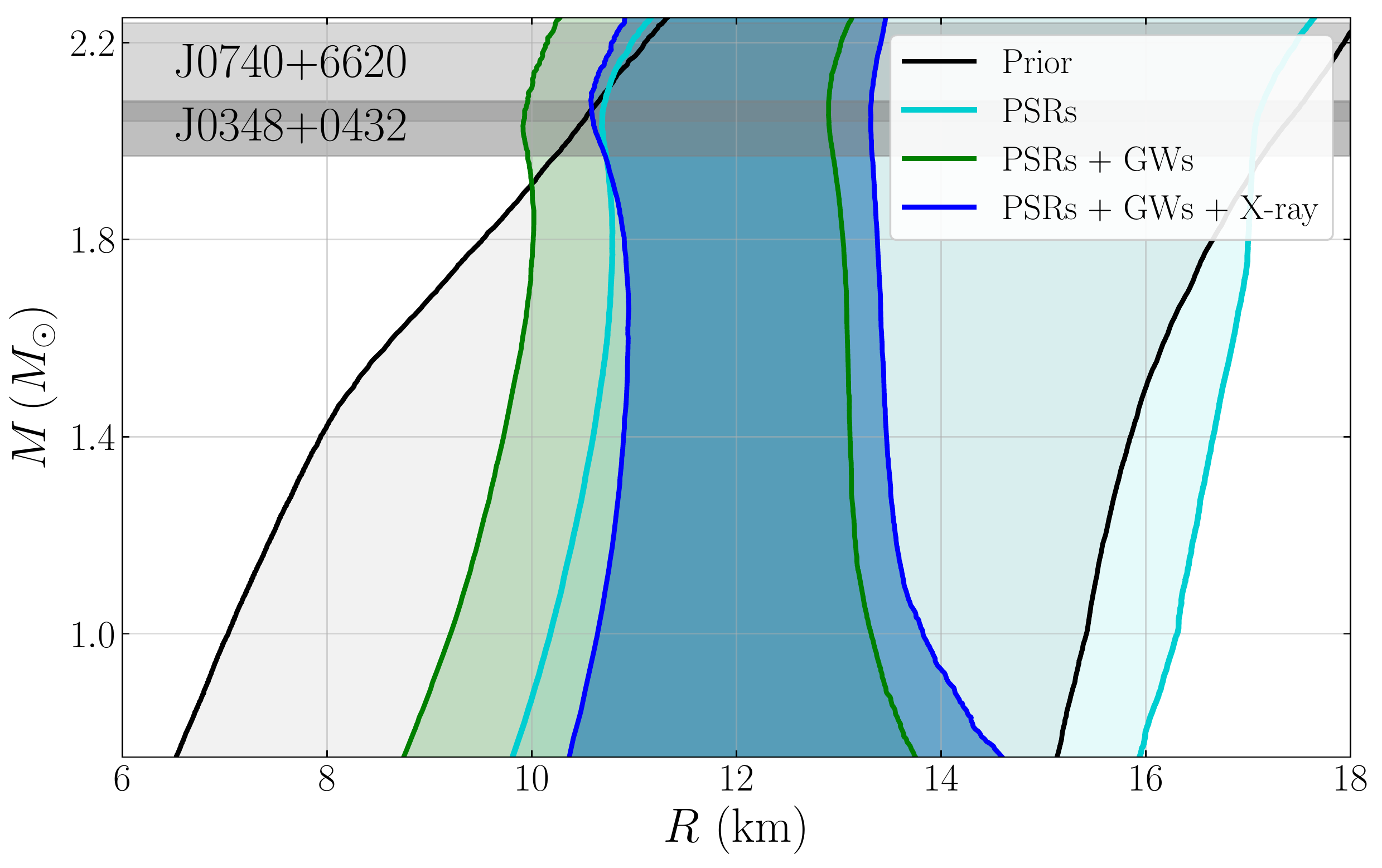}\\[-2ex]
    }
    \caption{
        Cumulative observational constraints on the EoS.
        We show 90\% symmetric credible intervals for the pressure as a function of density (left panel) and the radius as a function of mass (right panel). 
        Black contours denote the prior range, while turquoise contours correspond to the posterior when using only the heavy pulsar measurements.
        The other contours correspond to the posterior when also employing the GW (green) and NICER (blue) data.
        Vertical lines in the left panel denote multiples of the nuclear saturation density, while horizontal shaded regions in the right panel show the 68\% credible mass estimate for the two heaviest known pulsars.
    }
    \label{fig:EOScumulreal}
\end{figure*}

The main result of our analysis is a posterior process for the EoS that represents the cumulative impact of the astronomical observations. In Fig.~\ref{fig:EOScumulreal}, we plot this posterior process in the pressure-density (left panel) and mass-radius (right panel) planes.
The EoS constraints improve progressively as more data are added to our inference, starting from the prior (black lines) and incorporating the three heavy pulsars 
(turquoise), the GWs (green), and finally the NICER measurement (blue) in turn. As expected, the impact of the heavy pulsars is to prevent the NS pressure (or radius) from being too small (black to turquoise lines). The upper bounds on the tidal deformability from GW170817 and GW190425 translate into upper bounds on the NS pressure and radius (turquoise to green lines), while the NICER constraints serve to rule out some of the smaller pressures and radii (green to blue lines). The cumulative EoS constraints correspond to $R_{1.4}=\AllR\,$km for the radius of a canonical $1.4\,\Msolar$ NS and $p(2\rho_\mathrm{nuc})=\AllPtwonuc\times\PtwonucCoef\,\mathrm{dyn}/\mathrm{cm}^2$ for the pressure at twice nuclear saturation density ($\rho_{\rm nuc} = 2.8\times 10^{14}~\mathrm{g}/\mathrm{cm}^3$).

Besides providing up-to-date constraints on the EoS, we make projections for future constraints that could be achieved with further observations over the next $\sim5$ years. 
We consider additional detections of GW signals from BNS systems during upcoming LIGO-Virgo observing runs~\cite{Prospects2018} and hypothetical simultaneous mass-radius measurements for the announced NICER targets \cite{Guillot:2019vqp}. 
We simulate the likelihoods as multivariate Gaussian distributions with uncertainties inspired by GW170817 (rescaled by SNR) and J0030+0451, respectively.
LIGO and Virgo are expected to detect up to $\sim 60$ BNSs during their fourth observing run (c.~2022) \cite{2018arXiv181112907T}. Assuming four of those signals have SNR $>20$, we show that the NS radius uncertainty will decrease by $\sim 30\%$ from its current level.
During the fifth observing run (c.~2025), combined GW and NICER constraints can lead to ${\mathcal{O}(1)}\,$km uncertainty in $R_{1.4}$.
Moreover, we investigate the impact of a refined mass measurement for J0740+6620, or the discovery of an even heavier pulsar, and find that neither scenario would appreciably improve our knowledge of the EoS.
Similarly, we find that a possible first pulsar-timing measurement of the MoI of PSR J0737$-$3039, the double pulsar \cite{LyneBurgay2004,BurgayDAmico2003}, could help decrease the uncertainty of $R_{1.4}$ by $\sim 27\%$ relative to today---but that level of precision will be superseded by GWs from the fifth LIGO-Virgo observing run.

Indeed, while an individual X-ray observation with precision comparable to J0030+0451 typically provides better constraints on the canonical NS radius or the pressure at twice nuclear saturation density than a given GW observation, a greater number of BNS detections are expected over our $\sim 5$-year horizon, leading to projected constraints of \result{$\sim$10\%} uncertainty in $R_{1.4}$ and $\sim\MockDesignNICERMOIPtwonucRelError$ uncertainty in $p(2\rho_{\rm nuc})$ that are ultimately dominated by the GWs.

The rest of the paper is organized as follows.
In Sec.~\ref{sec:data} we provide details about the astronomical datasets we use.
In Sec.~\ref{sec:methods} we describe our methodology, namely the nonparametric EoS representation and the framework to combine multiple observations.
In Sec.~\ref{sec:EoScurrent} we present our current best constraints on the NS EoS.
In Sec.~\ref{sec:EoSfuture} we discuss potential future constraints from upcoming observations.
Finally, we conclude in Sec.~\ref{sec:conclusions}.

\section{Data}
\label{sec:data}

Our updated constraints on the EoS are based on three kinds of observations: radio surveys of heavy pulsars, GW signals from BNS coalescences, and X-ray emission from pulsar hot spots.
In this section we describe each dataset and discuss how it informs the EoS of supranuclear matter.
Tables~\ref{tab:Maxmassdata},~\ref{tab:BNSdata}, and~\ref{tab:NICERdata} summarize the data.
The observational data ($d$) are often reported as a finite set of posterior samples.
By reweighting the posterior samples to account for any nontrivial priors, we follow \cite{Landry:2018prl, Essick:2019ldf} and represent the likelihoods with optimized Gaussian kernel density estimates.
That is, given discrete samples of parameters $\theta$ drawn from the posterior probability distribution $p(\theta|d)$, we model the likelihood $p(d|\theta)$ at an arbitrary point in parameter space up to an overall normalization with a weighted kernel density estimate, assigning weights to each discrete posterior sample equal to the inverse of their prior probability $p(\theta)$.
Our kernel density estimate's bandwidth is chosen to maximize a cross-validation likelihood based on these weighted samples, as explained in Appendix B of~\cite{Essick:2019ldf}.

\subsection{Radio observations of massive pulsars}
\label{sec:heavy pulsars}

EoS models predict an absolute maximum mass for nonrotating NSs, the value of which is sensitive to the high-density behavior of the EoS~\cite{Lattimer_2005}.
The existence of NSs with masses  $\gtrsim2\,\Msolar$ suggests that the EoS is relatively stiff at high densities, and readily rules out EoS models that fail to support such massive stars.
Our analysis incorporates the masses of three such heavy pulsars (Table~\ref{tab:Maxmassdata}).
Each of these pulsars is in a binary system with a white dwarf companion, and the estimate of its mass was obtained either through a measurement of the Shapiro time delay (in the case of J1614$-$2230 and J0740+6620) or through spectral observations of the companion (in the case of J0348+0432).
We approximate the likelihood of the mass for each pulsar as a Gaussian that reproduces the reported median and uncertainty.

\begin{table}[]
    \begin{center}
        \begin{tabular}{l @{\quad}c}
            \hline \hline
            &\\[-2ex]
             PSR & $m$ $[\Msolar]$    \\[0.5ex]
              \hline 
            &\\[-2ex]
             J1614$-$2230~\cite{Demorest:2010bx,Fonseca:2016tux} &  $1.928^{+0.017}_{-0.017}$     \\[0.5ex]
             J0348+0432~\cite{Antoniadis:2013pzd} &  $2.01^{+0.04}_{-0.04}$     \\[0.5ex]
             J0740+6620~\cite{Cromartie:2019kug} & $ 2.14^{+0.10}_{-0.09}$     \\[0.5ex]
            \hline \hline  
        \end{tabular}
    \end{center}
    \caption{
        Summary of the heavy pulsar mass measurements we employ in this work.
        We quote the median and uncertainties ($68\%$ credible level) for the mass $m$ of each pulsar. The mass measurement we use for J1614$-$2230 has been superseded by $1.908^{+0.016}_{-0.016}\,\Msolar$ \cite{ArzoumanianBrazier2018}, but we do not expect this 1$\sigma$-level change in the median to affect our results.
    }
    \label{tab:Maxmassdata}
\end{table} 

\subsection{Binary neutron star coalescences via gravitational waves}
\label{sec:GW}

During the late stages of the inspiral of  coalescing NSs, the finite size of the stars gives rise to tidal interactions that affect the evolution of the binary system.
The tidal field produced by each binary component induces a quadrupole moment on the companion star, resulting in enhanced emission of gravitational radiation and a slight boost to their relative acceleration; the inspiral phase is sped up.
The magnitude of the induced quadrupole moment is related to the NSs' internal structure, with larger stars being less compact and thus more easily deformable under the influence of an external field of a given amplitude.
The effect is quantified through the dimensionless tidal deformability of each star $\Lambda_i$, $i\in \{1,2\}$, defined as the ratio of the induced quadrupole moment to the external perturbing tidal field \cite{Hinderer:2007mb}.
The tidal deformabilities can be directly constrained from analysis of the GW signal as they affect its phase evolution \cite{Flanagan:2007ix,Hinderer:2009ca}.

The advanced LIGO and Virgo detectors have so far detected GWs from the coalescence of two BNSs, GW170817 and GW190425. 
Analysis of each event yielded a multidimensional posterior distribution for the binary parameters, most notably the component masses $m_1, m_2$ and tidal deformabilities $\Lambda_1, \Lambda_2$. 
Table~\ref{tab:BNSdata} summarizes some of the relevant properties of each event.
We quote the chirp mass ${\cal{M}}\equiv (m_1m_2)^{3/5}/(m_1+m_2)^{1/5}$, the mass ratio $q\equiv m_2/m_1$, and the tidal parameter $\tilde{\Lambda}$~\cite{Favata:2013rwa}.
The latter, a particular mass-weighted combination of $\Lambda_1$ and $\Lambda_2$, is the best measured tidal parameter for GW170817 and GW190425, and the only tidal parameter expected to be measurable with current detector sensitivities~\cite{Wade:2014vqa}.
In our analysis we use the publicly available posterior samples from~\cite{170817samples} and~\cite{190425samples}. The posteriors are reported with respect to a prior that is uniform in the tidal deformabilities $\Lambda_1, \Lambda_2$ and the detector-frame component masses, subject to $m_2 \leq m_1$ \cite{Abbott:2018wiz,Abbott:2020uma}.

\begin{table}[]
    \begin{center}
        \begin{tabular}{l @{\quad}c @{\quad}c @{\quad}c@{\quad}}
            \hline \hline
            &&&\\[-2ex]
             BNS & ${\cal{M}}$ $[\Msolar]$&q & $\tilde{\Lambda}$   \\[0.5ex]
              \hline 
            &&&\\[-2ex]
             GW170817~\cite{TheLIGOScientific:2017qsa,Abbott:2018wiz}& $1.186^{+0.001}_{-0.001}$ & $(0.73,1.00)$ & $300^{+500}_{-190}$   \\[0.5ex]
             GW190425~\cite{Abbott:2020uma} & $1.44^{+0.02}_{-0.02}$ & $(0.8,1.0)$ & $\lesssim 600$     \\[0.5ex]
            \hline \hline 
        \end{tabular}
    \end{center}
    \caption{
        Summary of the BNS data we employ in this work.
        We quote the median (where applicable) and uncertainties ($90\%$ credible level) for the chirp mass ${\cal{M}}$, mass ratio $q$, and the tidal parameter $\tilde{\Lambda}$.
        We note that applying our nonparametric EoS priors can change these posterior credible regions somewhat, particularly for $\tilde{\Lambda}$~\cite{Landry:2018prl, Essick:2019ldf}, as the observables are correlated through the EoS.
    }
    \label{tab:BNSdata}
\end{table} 

\subsection{X-ray light curves from pulsar hot spots}
\label{sec:X-ray}

Precise modeling of the X-ray emission profile of hot spots on the surface of a rotating NS can yield an estimate of the stellar compactness, $C = G m/c^2 R$. The hot spots are small regions heated by cascades of particles from pair production in magnetosphere plasma gaps \cite{Bilous:2019knh}. The light curve of thermal X-rays from the heated regions varies in amplitude with the rotational period, and the widths of its peaks and the depth of its troughs are correlated with $C$, as greater lensing of a hot spot's emission means that its period of visibility increases at the expense of the totality of its eclipses. Relativistic Doppler and aberration effects in the light curve scale with the size of the object and its rotation rate, and can therefore be used to measure the stellar radius once the compactness is determined \cite{Bogdanov:2019qjb}. (In certain cases, e.g. for rapid rotation, these effects can in fact be large enough to constrain the radius more tightly than the compactness.)

Results from the pulse profile modeling of J0030+0451 were recently announced in two independent analyses~\cite{Miller:2019cac,Riley:2019yda}, and are summarized in Table~\ref{tab:NICERdata}.
The two analyses conclude that the observed pulse waveform is consistent with the emission from two or three noncircular hot spots, and they obtain samples from the multidimensional posterior distribution for the source parameters~\cite{Miller:2019cac,Riley:2019yda}. 
Overall, four separate hot spot models have been shown to reproduce the data accurately and produce broadly consistent results for the pulsar mass and radius (albeit with different priors on the radius): the 2- and 3-spot models from~\cite{Miller:2019cac} and the ST+PST and ST+CST models from~\cite{Riley:2019yda}.
The prior is uniform in mass $m$ and compactness $C$ in~\cite{Miller:2019cac}, and uniform in $m$ and $R$ in~\cite{Riley:2019yda}.
Since the posterior distributions obtained in~\cite{Riley:2019yda} rail against their mass and/or radius prior bounds, we use the 3-spot model from~\cite{Miller:2019cac} for our main results.
Nonetheless, as an alternative we also consider the model with the least similar mass-radius likelihood, ST+PST. In Appendix~\ref{sec:Riley}, we show that it leads to constraints on the EoS and on NS observables that are consistent with our primary analysis within statistical error. In other words, the systematic differences in the inferred mass and radius due to the assumed hot spot geometry appear to be small enough that they do not substantially affect the EoS information conveyed by the observation.
We emphasize that any differences between our main results and those in Appendix~\ref{sec:Riley} are not due to the different priors used in~\cite{Miller:2019cac,Riley:2019yda} as the priors have been removed following the hierarchical formalism laid out in Sec.~\ref{section:eos inference}.

\begin{table}[]
    \begin{center}
        \begin{tabular}{l @{\quad}c @{\quad}c}
            \hline \hline 
            &&\\[-2ex]
             PSR & $m$ $[\Msolar]$ & $R$ [km]    \\[0.5ex]
              \hline 
            &&\\[-2ex]
             J0030+0451, 2-spot~\cite{Miller:2019cac} &$1.44^{+0.19}_{-0.16}$& $13.27^{+1.41}_{-1.49}$     \\[0.5ex]
             \textbf{J0030+0451, 3-spot}~\cite{Miller:2019cac} &$\boldsymbol{1.44^{+0.15}_{-0.14}}$& $\boldsymbol{13.01^{+1.36}_{-1.06}}$     \\[0.5ex]
             \textbf{J0030+0451, ST+PST}~\cite{Riley:2019yda} &$\boldsymbol{1.34^{+0.16}_{-0.15}}$& $\boldsymbol{12.71^{+1.27}_{-1.18}}$     \\[0.5ex]
             J0030+0451, ST+CST~\cite{Riley:2019yda} &$1.43^{+0.19}_{-0.19}$& $13.86^{+1.34}_{-1.39}$     \\[0.5ex]
            \hline \hline  
        \end{tabular}
    \end{center}
    \caption{
        Summary of the NICER data; the models we use are shown in boldface.
        We quote the median and uncertainties ($68\%$ credible level) for the mass $m$ and the radius $R$ of the pulsar as computed from the released samples~\cite{miller_m_c_2019_3473466,riley_thomas_e_2019_3386449}.
        The intervals obtained with the ST+PST and ST+CST models are subject to prior bound cutoffs.
        Moreover, the radius posteriors are obtained under different priors between the two independent studies.
    }
    \label{tab:NICERdata}
\end{table} 

\section{Methodology}
\label{sec:methods}

\subsection{Nonparametric EoS inference}

We briefly review the Gaussian-process based nonparametric inference scheme developed in~\cite{Landry:2018prl, Essick:2019ldf}.
By assembling mixture models of Gaussian processes conditioned on tabulated EoSs from the literature, we construct generative models for synthetic EoSs and then compare them against the input data.
Gaussian processes support more model freedom than parametrized models of the EoS in that they do not prescribe a specific functional form for the EoS \textit{a priori}.
Any parametrization of the unknown NS EoS with a finite number of parameters is necessarily a lossy representation because the priors for the EoS parameters, regardless of their form or extent, assign exactly zero probability to any function that is not a member of the parametrized family.
In contrast, our nonparametric approach assigns a nonzero probability to all causal and thermodynamically stable EoSs.
In this way, our nonparametric inference is guaranteed to converge to the true EoS in the limit of infinite observations, assuming sufficient Monte Carlo sampling, and is not subject to the kind of modeling systematics inherent in parametrized inferences.

Although we condition our Gaussian processes on a set of theoretical proposals from the literature (see~\cite{Essick:2019ldf} for a complete list), our \emph{model-agnostic} prior depends on this set only weakly, generating synthetic EoSs that explore the entire space of plausible EoSs that are both causal and thermodynamically stable.
In this way, the \emph{model-agnostic} prior allows us to explore EoS models that differ significantly from those published in the literature, thereby reducing the impact of the precise choice of theoretical models used to construct our priors.
Indeed, our \emph{model-agnostic} priors are drastically less informative than the \emph{model-informed} priors from~\cite{Essick:2019ldf} that closely emulate only the behavior of the EoSs upon which they were conditioned.
Nonetheless, different choices for the input candidate EoSs, and different assumptions about the covariance kernels within the Gaussian process, would in general assign different prior probabilities to the synthetic EoSs (see~\cite{Essick:2019ldf}).
Like any EoS inference based on limited observational data, the posteriors we obtain are not completely independent of the EoS prior.

In this paper, we use the composition-marginalized \emph{agnostic} priors developed in~\cite{Essick:2019ldf}, which include generative models conditioned separately on EoSs that contain only hadronic matter ($pne\mu$), that contain hadronic and hyperonic matter ($pne\mu Y$), and that contain hadronic and quark (and possibly hyperonic) matter [$pne\mu(Y)Q$].
However, unlike~\cite{Landry:2018prl, Essick:2019ldf}, we do not require the synthetic EoSs generated by these priors to support stars of at least $1.93\,\Msolar$ \textit{a priori}.
Instead, knowledge of massive pulsars is incorporated via likelihood distributions, as described in Sec.~\ref{sec:heavy pulsars}; the massive pulsar data are included on an equal footing with all other astrophysical observations, such as coalescences seen in GWs (Sec.~\ref{sec:GW}), X-ray pulsations (Sec.~\ref{sec:X-ray}), and possible MoI measurements. 

\subsection{EoS inference with multiple observations}
\label{section:eos inference}

Accurate EoS inference from multiple observations requires us not only to model the EoS of NSs, but also to account for the properties of the population of detected sources as a whole and the selection effects inherent in our observations~\cite{Wysocki:2020myz}.
For example, BNSs detected through GWs depend on the mass distribution and rate of coalescing NSs, i.e. a population model, as well as the fact that GW detectors are more sensitive to heavy systems, i.e. selection effects.
Different datasets may represent different populations (e.g.~NSs in binaries vs.~in isolation) and may involve different selection effects.
In the current work, we assume flat mass prior distributions for all populations of events. We investigate how well the EoS can be constrained under that assumption with current data, as well as with a plausible set of hypothetical future detections.
While this may introduce small biases in the inferred EoS---the true astrophysical mass distribution is unlikely to be flat---these are expected to be smaller than the statistical uncertainty achievable in the near future.
Indeed, \cite{Wysocki:2020myz} shows that the wrong choice of population model only begins to seriously bias the inferred EoS after $\mathcal{O}(25)$ observations.
Nonetheless, a full analysis of both the underlying mass distributions and the EoS will be needed to avoid such systematics in the future.

Given the mass distributions we assume, we ignore selection effects for all practical purposes as these only influence the inference of the population model.
Nonetheless, in this section we describe a complete formalism that includes the effects of population models since they will become essential as more sources are detected.

Consider four classes of astronomical data that can be used to constrain the EoS: $d=\{\dgw,\ \dnicer,\ \dm,\ \dmoi\}$ denoting GW detections, NICER X-ray observations, heavy pulsar mass measurements, and MoI measurements, respectively.
Astronomical sources contributing to each class are described by certain population parameters $\lambda=\{\Lgw,\ \Lnicer,\ \Lm,\ \Lmoi\}$, which could, for example, describe the rate or distribution of coalescing BNSs in the Universe, or the mass distribution of NICER pulsars.
Each class can contain more than one source (e.g.~GW170817 and GW190425 for GW).
The posterior probability for a single EoS, $\varepsilon(p)$, is given by
\begin{align}
    P(\varepsilon|d) = \frac{P(d|\varepsilon) P(\varepsilon)}{P(d)} = \frac{P(\varepsilon) \prod_i  P(d_i|\varepsilon)}{P(d)} ,
\end{align}
where $i$ enumerates the four data classes. The form of the likelihood $P(d_i|\varepsilon)$ depends on the data class and is given by the product of the likelihoods for individual signals, the forms of which are discussed below.
Intuitively, the final expressions reduce to integrals of macroscopic observables determined by the EoS, such as $\Lambda(m)$, $R(m)$, and $I(m)$, over the likelihood function from observations of the relevant parameters.

\subsubsection{GW detections}

For GW signals, the relevant observational parameters are the component masses $m_1,\ m_2$, the tidal deformabilities $\Lambda_1,\ \Lambda_2$, and the population parameters $\Lgw$.
In this case
\begin{align}
    P(\dgw|\varepsilon) &= \int dm_1 \int dm_2 \int d \Lambda_1 d \Lambda_2 \int d \Lgw \, P(\Lgw) \nn \\
                        &\quad\quad \times P(m_1,m_2,\Lambda_1,\Lambda_2|\varepsilon,\Lgw)\nn \\
                        &\quad\quad \times \frac{P(\dgw|m_1,m_2,\Lambda_1,\Lambda_2)}{\beta(\Lgw)} , \label{GWlikeL}
\end{align}
where $P(\Lgw)$ is the prior over population parameters.
The denominator $$\beta(\Lgw) \equiv \int d m_1 dm_1 P_\mathrm{det}(m_1, m_2) P(m_1,m_2|\Lgw),$$ where $P_\mathrm{det}(m_1, m_2)$ is the probability of a system with masses $m_1,\ m_2$ being detected by the network of GW detectors, encompasses the selection effects of the survey after marginalizing over the unknown overall rate $R_\mathrm{GW}$ of BNSs with a prior $P(R_\mathrm{GW}) \sim 1/R_\mathrm{GW}$.
See~\cite{Mandel:2018mve} for a full derivation.\footnote{In general, $\beta(\Lgw)$ could depend on $\varepsilon$ as well, but a system's detectability is dominated by its masses and not $\varepsilon$, so we approximate it with only the dependence on the mass distribution.}

Since, for a given EoS $\varepsilon$, the tidal deformability is a function of the mass, Eq.~\eqref{GWlikeL} can be further simplified to
\begin{align}
    P(\dgw|\varepsilon) 
                        &= \int dm_1 \int dm_2 \int d \Lgw \, P(\Lgw) \nn \\
                        &\quad\quad \times P(m_1, m_2|\varepsilon,\Lgw) \nn \\ 
                        &\quad\quad \times \frac{P(\dgw|m_1,m_2,\Lambda_1(m_1,\varepsilon),\Lambda_2(m_2,\varepsilon))}{\beta(\Lgw)}  \label{GWlikeLPop}
\end{align}
by writing \pagebreak

\begin{align}
    P(m_1,m_2,\Lambda_1,\Lambda_2|\varepsilon,\Lgw) = & P(m_1,m_2|\Lgw) \nn \\
                                                      & \, \times \delta(\Lambda_1-\Lambda_1(m,\varepsilon)) \nn \\
                                                      & \, \times \delta(\Lambda_2-\Lambda_2(m,\varepsilon)) .
\end{align}
We take the domain of the function $\Lambda(m,\varepsilon)$ prescribed by the EoS to be all $m > 0$, with $\Lambda(m,\varepsilon) = 0$ (the expected BH value~\cite{Binnington:2009bb}) for $m > M_{\rm max}(\varepsilon)$.
Equation~\eqref{GWlikeLPop}, then, infers the properties of the NS EoS from GW observations while simultaneously marginalizing over the rates and mass distribution of coalescing low-mass compact objects.
Indeed, given the close relation between the inferred NS masses and EoS inference it is not surprising that incorporating the wrong population model for a large number of detections can lead to biased EoS inference~\cite{Agathos:2015uaa,Wysocki:2020myz}.
However, because of the small number of detected signals for this study, we choose to fix the population model such that NS masses are drawn from a uniform distribution with $m \geq 0.5\,\Msolar$,\footnote{In practice, the upper bound of the assumed NS mass prior is limited by the observational likelihood's domain of support. The same principle applies to the choice of mass range for the NICER observations below.} assuming BHs cannot exist below $M_{\rm max}(\varepsilon)$.
We leave full population marginalization to future work.
In this case, the selection term $\beta(\Lgw)$ is constant and
\begin{align}
    P(\dgw|\varepsilon) \sim & \int dm_1 \int dm_2  \, P(m_1, m_2) \nn \\
    &\;\; \times P(\dgw|m_1,m_2,\Lambda_1(m_1,\varepsilon),\Lambda_2(m_2,\varepsilon)), \label{GWlikeLfinal}
\end{align}
where we have dropped the explicit dependence on $\Lgw$ for notational simplicity and have simplified $P(m_1, m_2|\varepsilon)=P(m_1, m_2)$.

However, in general, the prior on component masses $P(m_1, m_2| \Lgw)$ may have support above the maximum mass $\Mmax$ predicted by a given EoS.
By default, we do not assume that each observation is of a BNS; i.e. we admit the possibility that a given component of the source may be a BH, and this is the case described by Eq.~\eqref{GWlikeLfinal}.

If instead we knew in advance that a particular coalescence was definitely a BNS (e.g.~via observation of an electromagnetic counterpart), then that knowledge could be incorporated into our inference as

\begin{align}
    P(\dgw|\varepsilon,\mathrm{NS}) \sim & \int dm_1 \int dm_2  \, P(m_1, m_2|\varepsilon,\mathrm{NS}) \nn \\
    &\;\; \times P(\dgw|m_1,m_2,\Lambda_1(m_1,\varepsilon),\Lambda_2(m_2,\varepsilon)),
\end{align}
where

\begin{multline}
    P(m_1, m_2|\varepsilon, \mathrm{NS}) = \\
    P(m_1, m_2) \frac{\Theta(m_1\leq \Mmax)\Theta(m_2\leq m_1)}{\int\limits^{\Mmax} d{m_1}' \int\limits^{m_1} d{m_2}' P({m_1}', {m_2}')} .
\end{multline}

We emphasize that the explicit normalization of the mass prior above is crucial.
The normalization acts as an Occam factor that prefers EoSs with $\Mmax$ only slightly larger than the largest observed mass and disfavors EoSs with larger $\Mmax$, as appropriate if we assume the maximum observed mass is limited by the EoS; see Appendix~\ref{sec:Occam factors} for more details.
In contrast, Eq.~\eqref{GWlikeLfinal} does not penalize EoSs based on their value of $\Mmax$~\cite{Chatziioannou:2015uea} besides the intrinsic correlations with $\Lambda$ within the GW likelihood.
The inferences performed in this work do not assume that either GW170817 or GW190425 were known \textit{a priori} to be BNSs, and we therefore use Eq.~\eqref{GWlikeLfinal} with priors uniform in the component masses.

\subsubsection{NICER observations}

For NICER observations, the relevant measurement concerns the mass $m$ and radius $R$ of a single pulsar, and the population properties $\lambda_{\mathrm{X}}$ of the pulsars targeted by NICER.
Following the same steps as for GW observations, we obtain an expression for the NICER likelihood that marginalizes over population parameters and includes the relevant selection effects:
\begin{align}
    P(\dnicer|\varepsilon,\mathrm{NS})  =& \int dm \int d \Lnicer \, P(\Lnicer) \nn \\
    &\;\; \times P(m|\varepsilon,\mathrm{NS},\Lnicer) \frac{P(\dnicer|m,R(m;\varepsilon))}{\beta(\Lnicer)} \nn \\
    &\sim \int dm \, P(m|\varepsilon,\mathrm{NS}) P(\dnicer|m,R(m;\varepsilon)) . \label{NICERlikeLPop}
\end{align}
In the last expression we have again assumed a fixed population of pulsars observed by NICER with masses uniformly distributed above $0.5\,\Msolar$, dropping the implicit dependence on $\Lnicer$ for simplicity.

Here, the X-ray observations are predicated on the fact that the object is a NS and therefore we must account for that prior knowledge.
This is why we retain the dependence on $\varepsilon$ within
\begin{align}
    P(m|\varepsilon,\mathrm{NS},\Lnicer) = \frac{P(m|\Lnicer) \Theta(m\leq \Mmax)}{\int\limits^{\Mmax} dm' \, P(m'|\Lnicer)},
\end{align}
where we explicitly account for the prior normalization.
However, in our analysis, we assume that the population of NSs targeted by NICER observations have masses much lower than $\Mmax$ from any EoS that is compatible with the existence of massive pulsars.
Therefore, the assumed population described by $\Lnicer$ truncates at masses below $\Mmax$ for any relevant $\varepsilon$, and $p(m|\varepsilon,\mathrm{NS},\Lnicer)=p(m|\Lnicer)$ as the normalization is limited by $\Lnicer$ rather than $\Mmax$.

\subsubsection{Massive pulsars}

Regarding heavy pulsar observations, a common approach is to simply reject all EoSs that do not support masses above a predetermined threshold~\cite{Lackey:2014fwa,Carney:2018sdv,Abbott:2018exr}.
Here we instead follow the approach of~\cite{Miller:2019nzo} and others, marginalizing over the mass measurement by taking into account the measurement uncertainty.
Specifically,
\begin{align}
    P(\dm|\varepsilon,\mathrm{NS}) &= \int dm \int d \Lm \, P(m|\varepsilon,\mathrm{NS},\Lm) \frac{P(\dm|m)}{\beta(\Lm)} \nn \\
    &\sim \int dm \, P(m|\varepsilon,\mathrm{NS}) P(\dm|m) .
\end{align}
Like the NICER observations, mass measurements of pulsars assume the objects are NSs and therefore
\begin{equation}
    P(m|\varepsilon,\mathrm{NS},\Lm) = \frac{P(m|\Lm) \Theta(m\leq \Mmax)}{\int\limits^{\Mmax} dm' \, P(m'|\Lm)} .
\end{equation}
In this case, the explicit normalization term in the prior must be taken into account as the observed masses are close to the maximum masses predicted by EoSs in our prior.

We assume flat priors up to $\Mmax$ and Gaussian likelihoods for $p(\dm|m)$; $p(\dm|\varepsilon,\mathrm{NS})$ is thus a sigmoid (error function) disfavoring $\varepsilon$ with small $\Mmax$. Its width is determined by the measurement uncertainty in the pulsar's mass.

\subsubsection{Moment of inertia measurements}

Finally, we treat the measurement of a NS's MoI. Although no such measurements currently exist, they have long been anticipated \cite{LyneBurgay2004,LattimerSchutz2005,KramerWex2009} and we consider them in the context of our projected constraints in Sec.~\ref{sec:EoSfuture}. For measurements relating the mass $m$ and MoI $I$ from a population $\Lmoi$, we obtain
\begin{align}
    P(\dmoi|\varepsilon,\mathrm{NS}) =& \int dm \int d \Lmoi \, P(\Lmoi) \nn \\
    &\;\; \times P(m|\varepsilon,\mathrm{NS},\Lmoi) \frac{P(\dmoi|m,I(m;\varepsilon))}{\beta(\Lmoi)} \nn \\
    &\sim \int dm \, P(m|\varepsilon,\mathrm{NS}) P(\dmoi|m,I(m;\varepsilon)). \label{MOIlikeLPop}
\end{align}
MoI measurements assume the object is a NS and we therefore retain the dependence on the EoS within the mass prior:
\begin{equation}
    P(m|\varepsilon,\mathrm{NS},\Lmoi) = \frac{P(m|\Lmoi) \Theta(m\leq \Mmax)}{\int\limits^{\Mmax} dm \, P(m|\Lmoi)} .
\end{equation}
As with NICER data, we assume $\Mmax$ is larger than any mass allowed by $\Lmoi$ so that $p(m|\varepsilon,\mathrm{NS},\Lmoi)=p(m|\Lmoi)$ in practice.

\section{Current constraints on the EoS}
\label{sec:EoScurrent}

We provide updated constraints in the pressure-density and mass-radius planes in Sec.~\ref{sec:pressure density}, on the sound speed as a function of density in Sec.~\ref{sec:sound speed}, and on individual source properties based on the joint EoS inference in Sec.~\ref{sec:source properties}. We discuss implications for hybrid stars in Section~\ref{sec:hybrid stars}.

\subsection{Pressure-density and mass-radius relations}
\label{sec:pressure density}

Combining the existing heavy pulsar, GW, and NICER results according to the methodology described in the previous section, we obtain posteriors for the EoS as well as various NS properties.
Figure~\ref{fig:EOScumulreal} shows the cumulative posterior for the EoS in the pressure-density (left) and mass-radius (right) planes.
The posteriors are also shown event by event in Fig.~\ref{fig:EOSindivreal} of Appendix~\ref{sec:moreplots}.
To obtain each plot, we compute the 90\% symmetric credible intervals for the pressure (radius) for a fixed density (mass) and plot them as a function of the density (mass).
The black lines correspond to the \emph{model-agnostic} prior described in Sec.~\ref{sec:methods} A.

The turquoise, green, and blue posteriors are obtained by successively incorporating more observations.
The turquoise lines correspond to the posterior after taking into account the heavy pulsar measurements from Table~\ref{tab:Maxmassdata}.
Their main effect is to prevent the pressure from being too low at large densities, thereby disfavoring a large part of the lower end of the pressure-density prior.
In the mass-radius plane, the existence of heavy pulsars is similarly inconsistent with very small radii~\cite{Annala:2017llu}, 
while being relatively uninformative for large radii.
Though the turquoise line is obtained with all three heavy pulsars from Table~\ref{tab:Maxmassdata}, we find that the information they provide is primarily driven by J0348+0432~\cite{Antoniadis:2013pzd}, due to its large mass and small measurement error.
Indeed, J0740+6620~\cite{Cromartie:2019kug} may appear heavier and thus more constraining at first glance, but its large uncertainty makes it essentially consistent with J0348+0432.
In anticipation of a more precise mass measurement for J0740+6620, in the next section we explore the effect on the EoS constraints of a decreased measurement uncertainty for this pulsar.

The green lines and shaded region show the EoS posterior after incorporating both GW and heavy pulsar measurements.
Unsurprisingly, we find that the GW data disfavor large pressures and large radii, as they are more consistent with soft EoSs due to the small magnitude of the observed tidal deformations.
As already discussed in~\cite{Abbott:2020uma}, we also find that GW190425 offers very little information about the EoS given its low SNR; the GW information comes almost exclusively from GW170817.
This is also consistent with earlier results from~\cite{Lackey:2014fwa} that suggest that joint EoS constraints are primarily driven by only the loudest GW signals.
According to our analysis, GW and heavy pulsar data jointly imply a radius of $R_{1.4}= \PSRsGWsR$ km for a $1.4\,\Msolar$ NS, consistent with other studies, e.g.,~\cite{Abbott:2018exr,Landry:2018prl}.

Finally, the blue lines and shaded region correspond to the posterior using all three relevant datasets, including the recent NICER constraints on the mass and radius of J0030+0451.
We find that J0030+0451 disfavors the lowest pressures and radii compatible with GW observations, which are ultimately excluded at the 90\% confidence level.
This suggests that current GW observations favor slightly softer EoSs than J0030+0451.
Indeed the upper limit on the pressure (or radius) is mostly driven by the GW data, which place strong constraints on the EoS's stiffness.
On the other hand, the lower limit on the pressure (or radius) is determined jointly by all observations. 
Overall we find $R_{1.4}=\AllR$ km and $p(2\rho_{\rm nuc}) = \AllPtwonuc\times\PtwonucCoef\,\mathrm{dyn}/\mathrm{cm}^2$ at the 90\% credible level.
Table~\ref{tab:ResultsSummary} shows joint constraints on other observables, while Appendix~\ref{sec:moreplots} hosts additional results, including the individual-event (rather than cumulative) EoS posteriors and  correlations between pertinent parameters.

\subsection{Sound speed}
\label{sec:sound speed}

The speed of sound in NS matter gives an indication of the microscopic interactions that govern cold matter at supranuclear densities. At extremely high densities (not necessarily realized in NSs), the squared sound speed $c_s^2 =  c^2 \, dp/d\varepsilon$ is expected to approach the ultrarelativistic limit of $c^2/3$ from below, characteristic of asymptotically free quarks \cite{KurkelaRomatschke2010}.
It has been conjectured~\cite{PhysRevLett.114.031103} that the sound speed might also satisfy $c_s^2 < c^2/3$ throughout the whole NS, although there are now many nuclear-theoretic models for NS matter that violate this so-called conformal limit, e.g.~\cite{KojoPowell2015,BitaghsirFadafanKazemian2018,McLerranReddy2019}.
Evidence for $c_s^2 > c^2/3$ within NSs would imply that the NS matter is strongly interacting and nonconformal at the relevant densities~\cite{PhysRevLett.114.031103,Tews_2018SOUNDSPEED}.

\begin{figure}[]
    \parbox{\hsize}{
    \includegraphics[width=\hsize, clip=True]{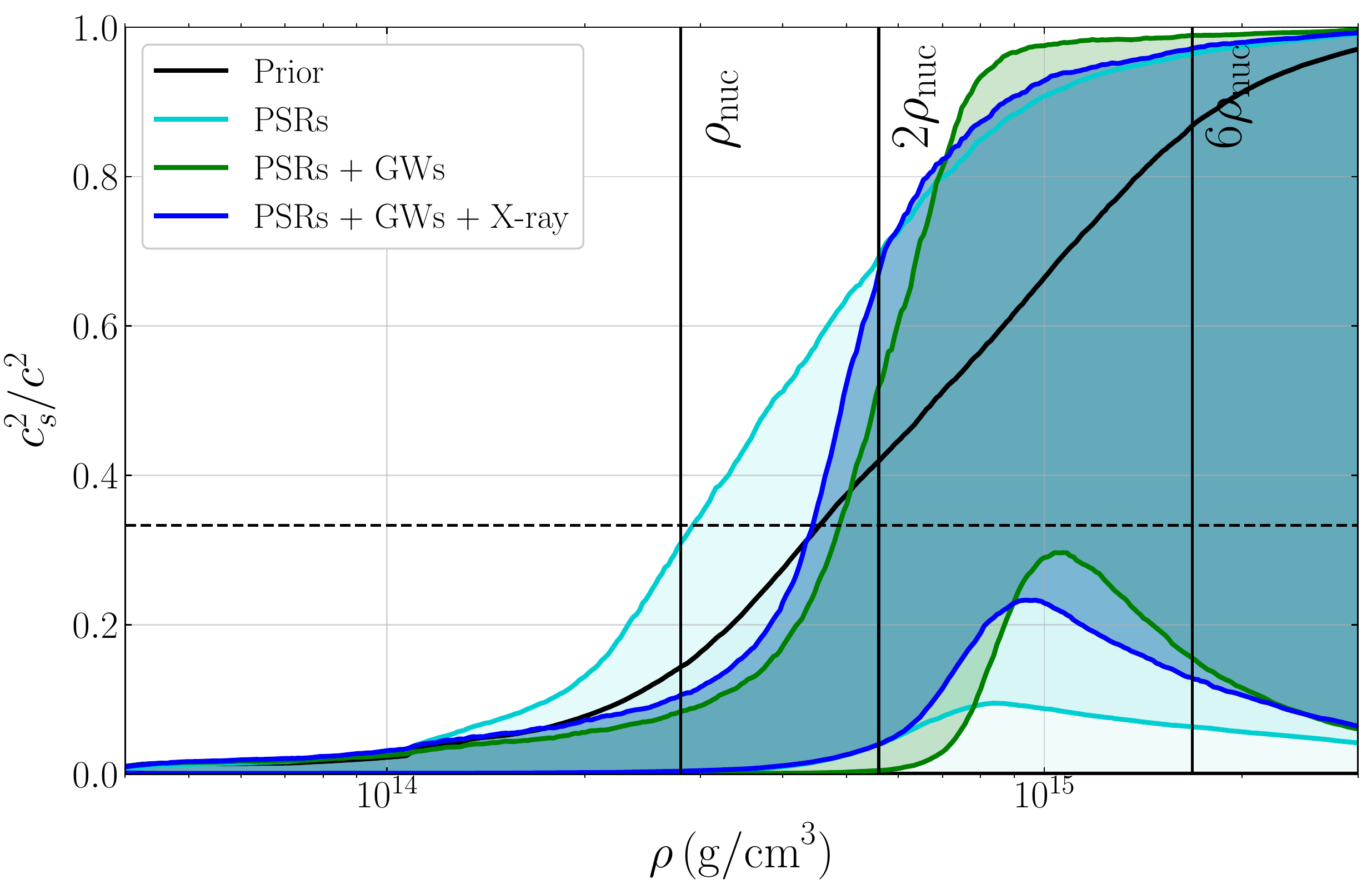}\\[-2ex]
    }
    \caption{
        Constraints on the sound speed in NS matter.
        Contours denote 90\% credible regions for the various combinations of astronomical data.
        The conjectured conformal bound $c_s^2/c^2 < 1/3$ is shown as the horizontal dashed line.
    }
    \label{fig:cs2}
\end{figure}

Figure~\ref{fig:cs2} shows the impact of different combinations of astrophysical datasets on the inferred sound speed.
The conformal limit of $c_s^2 \rightarrow c^2/3$ is also plotted for comparison.
Interestingly, all of the astrophysical datasets increase the support for sound speeds above the conformal limit relative to the prior.

The existence of $\sim 2\,\Msolar$ pulsars is known to favor sound speeds in excess of the conformal limit~\cite{PhysRevLett.114.031103,AlsingSilva2018,Tews_2018SOUNDSPEED}. We find here that GW and NICER data strengthen this preference, nearly ruling out the possibility that $c_s^2 < c^2/3$ at all densities. Reference~\cite{ReedHorowitz2020} arrived at a similar conclusion by studying the sound speeds of a discrete set of low-density EoSs matched to constant sound-speed extensions that are compatible with the observations of GW170817 and PSR J0740+6620.
The increased preference for $c_s^2 > c^2/3$ is driven to a large extent by the GW observations, which favor soft EoSs at low densities and therefore require a large sound speed above nuclear saturation density in order to support $\sim 2\,\Msolar$ pulsars.
The incorporation of NICER data slightly reduces the maximum inferred sound speed relative to the GWs, as the NICER observation prefers stiffer EoSs at low densities.

In Table~\ref{tab:ResultsSummary}, we report credible regions for the maximum sound speed attained inside NSs---i.e. at any density below the central density of the maximum-mass stellar configuration for each synthetic EoS.
Our prior already has significant support for a maximum sound speed above the conformal limit, and the incorporation of astrophysical data only strengthens this conclusion, consistently ruling out maximum sound speeds $\leq c/\sqrt{3}$ at 90\% credibility.
Although very suggestive, this does not imply, however, that the conformal limit is irreconcilably inconsistent with current data.
The table also reports the density and pressure at which the maximum $c_s^2$ is reached; we find that it typically occurs between \result{3.5--4.5} times nuclear saturation density, just above the central densities inferred for the components of GW170817~\cite{Essick:2019ldf} and PSR J0030+0451~\cite{Raaijmakers:2019qny}.
These constraints strongly suggest that NS matter is nonconformal and strongly coupled around $\sim4\,\rho_{\rm nuc}$.

\subsection{Source properties}
\label{sec:source properties}

\begin{figure}[]
    \parbox{\hsize}{
    \includegraphics[width=\hsize, clip=True]{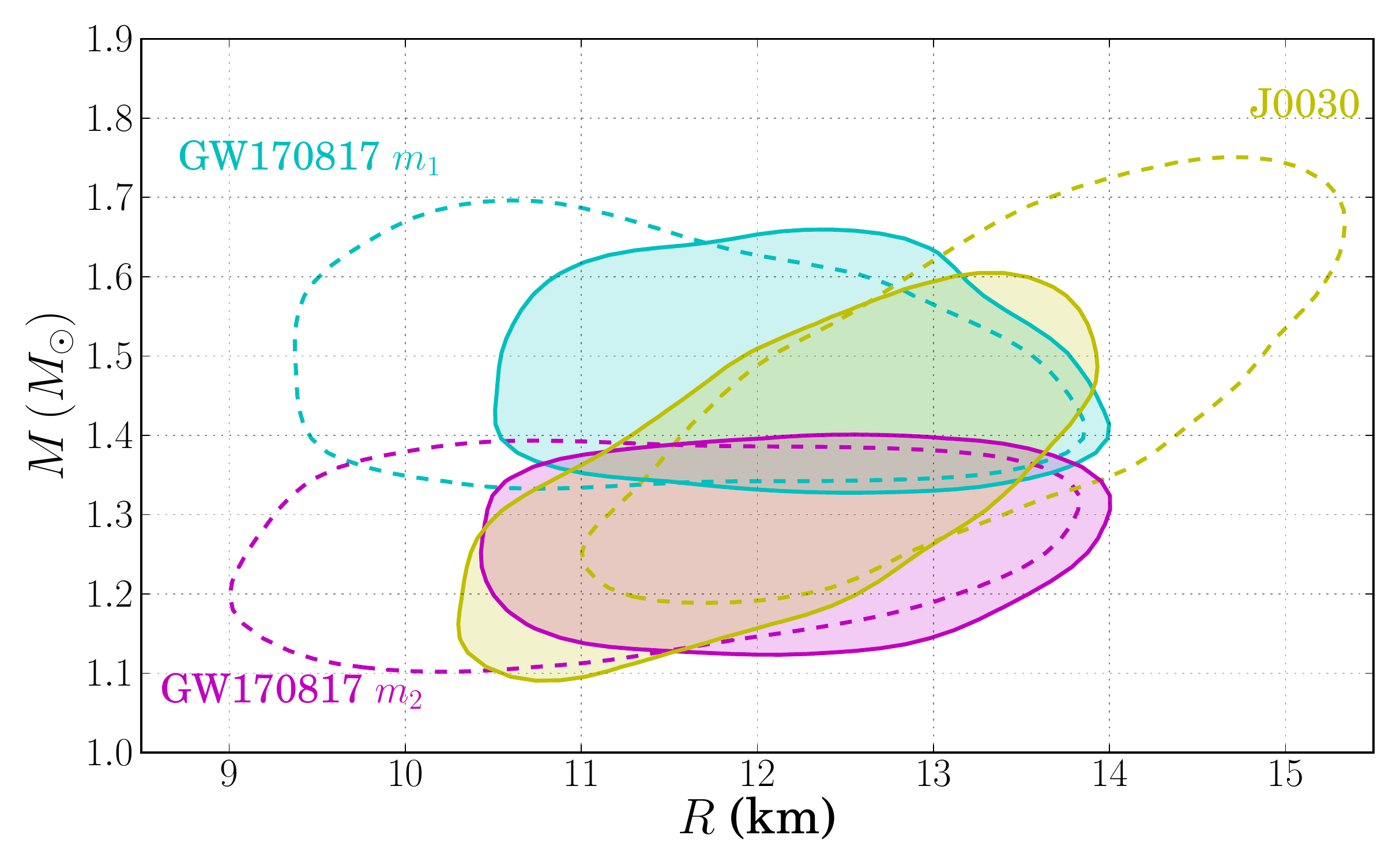}\\[-2ex]
    }
    \caption{
        Contours of the two-dimensional mass-radius posteriors for the two NSs in GW170817 and for J0030+0451.
        Dashed lines correspond to 90\% credible regions from analyses that only use each respective observation plus the heavy pulsars, while solid lines and shading correspond to 90\% credible regions from the joint analysis that employs all the astronomical data.
    }
    \label{fig:MR170817NICER}
\end{figure}

Figure~\ref{fig:MR170817NICER} shows the 90\% contours of the separate two-dimensional mass-radius posteriors for GW170817's binary components and J0030+0451.
The dashed lines correspond to the posterior when incorporating only data from the respective observation and the heavy pulsars.
The solid lines and shaded regions show the posterior after incorporating joint information from all the astronomical observations. The true $M$-$R$ relation must pass through each of the individual-event posteriors (though not necessarily simultaneously).
The plot confirms that the GW data systematically favor lower radii than the NICER data on their own, though there is significant overlap between the two.
The combined analysis favors the overlapping region around $12\,$km, which is consistent with both individual observations.
Given both the GW and the NICER data, plus the existence of heavy pulsars, we find that the radius of the primary GW170817 component is $\UpdatedOhEightRone\,$km, while the radius of J0030+0451 is $\UpdatedNicerR\,$km at the 90\% credible level. Similarly EoS-informed source properties for the secondary component of GW170817 and for the components of GW190425 are displayed in Table~\ref{tab:updated params}. The EoS-informed value for J0030+0451's radius could, in principle, help determine the preferred hostpot geometry by enabling a consistency test of the various models' radius predictions. 
At present, the statistical uncertainties in both the predicted and the measured $R$ values are too large for such a test to be informative, but it could be worthwhile for a future NICER target whose mass is known \textit{a priori}.

\subsection{Hybrid stars}
\label{sec:hybrid stars}

\begin{figure}[]
    \parbox{\hsize}{
    \includegraphics[width=\hsize]{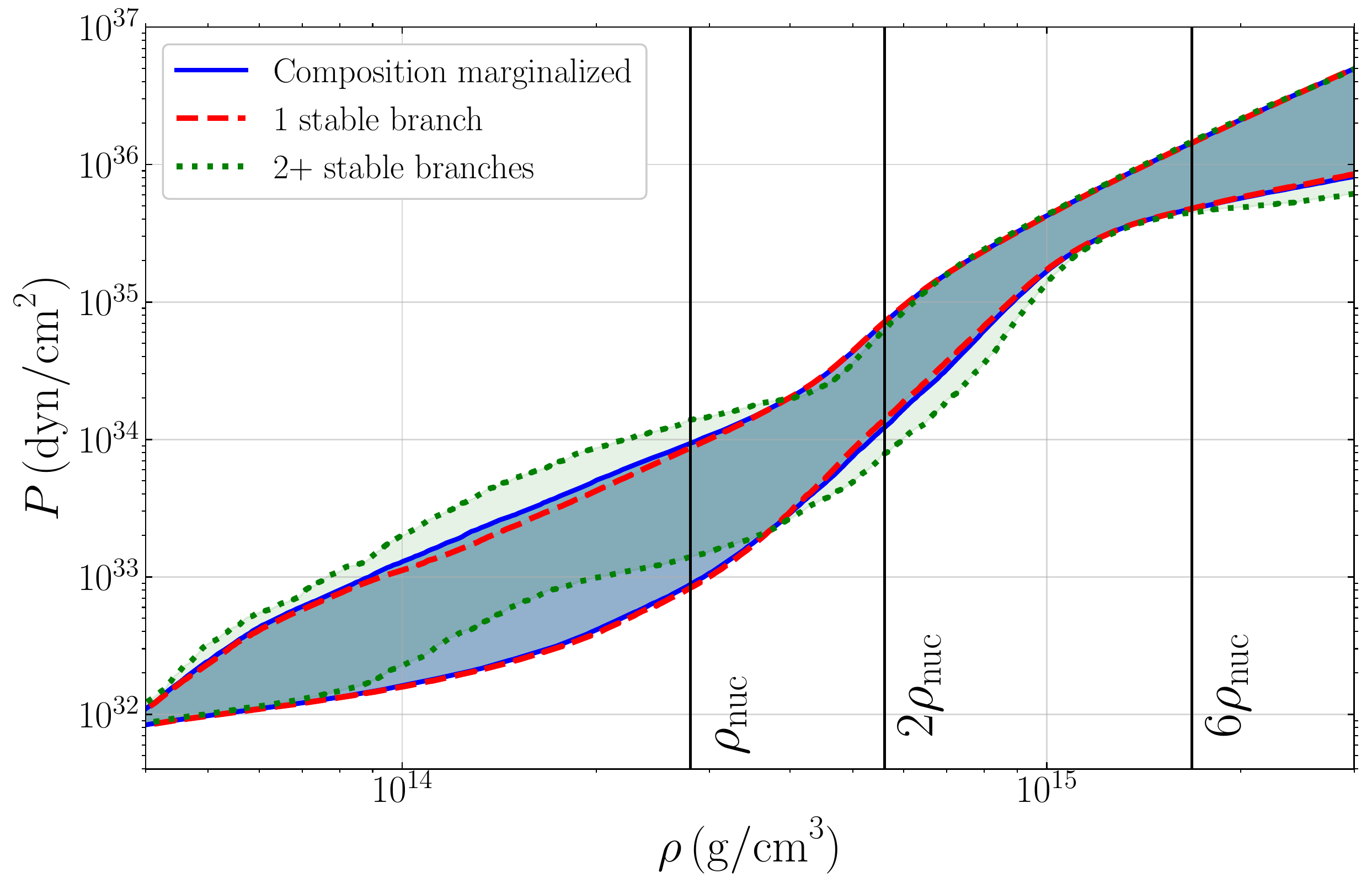}\\[-2ex]
    }
    \caption{
	Effect of EoS phenomenology.
	We show 90\% symmetric credible intervals for the EoS in the pressure-density plane. 
        The blue lines and shaded region correspond to the results obtained after marginalizing over the EoS phenomenology (identical to the blue lines in Fig.~\ref{fig:EOScumulreal}). Red dashed lines (green dotted lines) denote the pressure posterior for EoSs that contain only one stable branch (two or more stable branches) in their mass-radius relation. 
    }
    \label{fig:EOScumulrealPT}
\end{figure}

The analysis presented in Figs.~\ref{fig:EOScumulreal}-\ref{fig:MR170817NICER} is not restricted to any specific assumption about the composition of NSs, as it is based on EoSs that have been drawn, in equal numbers, from priors conditioned on hadronic, hyperonic and quark EoS models alike.
More importantly, the priors are \emph{agnostic} and generate novel behavior relative to the candidate EoSs upon which the Gaussian process was conditioned.
Figure~\ref{fig:EOScumulrealPT} breaks down the EoS constraints according to the different EoS phenomenologies in the prior.
Specifically, we condition the prior based on the number of stable branches in the $M$-$R$ relation, i.e.~based on whether the EoS supports a disconnected hybrid star branch. The presence of more than one stable branch indicates a strong phase transition in the EoS, though the converse is not true: a single continuous stable branch may also be consistent with a phase transition. The existence of hybrid stars with separate hadronic and quark phases has been proposed by many different authors \cite{Gerlach1968,Kampfer1981,GlendenningKettner2000,SchertlerGreiner2000,AlfordSedrakian2017,AlvarezCastilloBlaschke2017,LiYan2018,PaschalidisYagi2018,NandiChar2018,MontanaTolos2019,BurgioFigura2019}, and here we investigate whether hybrids remain compatible with the EoS constraints gleaned from astronomical observations.
 
We find that the EoS posteriors under the different assumptions about the NS phenomenology largely agree with each other at the 90\% credible level, implying that current observational results cannot distinguish between EoSs that support hybrid stars or not with high confidence. The Bayes factor comparing evidence for multiple stable branches to a single stable branch, assuming the existence of massive pulsars \textit{a priori}, is only
\begin{equation}
    B^{n>1}_{n=1} = \frac{p(d_\mathrm{GW}, d_\text{X}|d_\mathrm{PSR}, n>1)}{p(d_\mathrm{GW},d_\text{X}|d_\mathrm{PSR}, n=1)} = \BayesTwoOneBranchGivenPSR
\end{equation}
when both GW and X-ray data are used.
This is slightly smaller than what was reported in~\cite{Essick:2019ldf}, corresponding to a slightly larger preference \textit{a posteriori} for stiffer EoSs, which preferentially have only a single stable branch, when X-ray observations are included.\footnote{If we do not condition on the existence of massive pulsars \textit{a priori}, we find a Bayes factor of $\BayesTwoOneBranch$, due primarily to the fact that massive pulsars rule out many of the EoS with multiple stable branches in our prior, as they are too soft to support $2\,\Msolar$ stars.}
We obtain $R_{1.4}=\AllRonebranch\,$km and $\AllRtwobranch\,$km for EoSs with a single and multiple stable branches, respectively.

\begin{table*}[]
    \begin{center}
        \begin{tabular}{l @{\quad}r @{\quad}r @{\quad}r @{\quad}r @{\quad}r}
            \hline \hline
            Observable              & Prior           & PSRs           & PSRs+GWs      & PSRs+X-ray     & PSRs+GWs+X-ray \\
            \hline
            $\Mmax$ $[\Msolar]$    & \PriorMmax      & \PSRsMmax      & \PSRsGWsMmax      & \PSRsXrayMmax      & \AllMmax \\
            $R_{1.4}$ $[{\rm km}]$ & \PriorR         & \PSRsR         & \PSRsGWsR         & \PSRsXrayR         & \AllR \\
            $\Lambda_{1.4}$       & \PriorL         & \PSRsL         & \PSRsGWsL         & \PSRsXrayL         & \AllL \\
            $I_{1.4}$ $[10^{45}\,\mathrm{g}\,\mathrm{cm}^2]$
                                  & \PriorI         & \PSRsI         & \PSRsGWsI         & \PSRsXrayI         & \AllI \\
            $p(\rho_\mathrm{nuc})$ $[\PnucCoef{\rm dyn}/{\rm cm^2}]$       & \PriorPnuc      & \PSRsPnuc      & \PSRsGWsPnuc      & \PSRsXrayPnuc      & \AllPnuc \\
            $p(2\rho_\mathrm{nuc})$ $[\PtwonucCoef{\rm dyn}/{\rm cm^2}]$   & \PriorPtwonuc   & \PSRsPtwonuc   & \PSRsGWsPtwonuc   & \PSRsXrayPtwonuc   & \AllPtwonuc \\
            $p(6\rho_\mathrm{nuc})$ $[\PsixnucCoef{\rm dyn}/{\rm cm^2}]$   & \PriorPsixnuc   & \PSRsPsixnuc   & \PSRsGWsPsixnuc   & \PSRsXrayPsixnuc   & \AllPsixnuc \\
            $\max c_s^2/c^2$ & \PriorMaxCs & \PSRsMaxCs & \PSRsGWsMaxCs & \PSRsXrayMaxCs & \AllMaxCs \\
            $\rho\left(\max c_s^2/c^2\right)$ $[10^{15}\mathrm{g}/\mathrm{cm}^3]$ & \PriorRhoMaxCs & \PSRsRhoMaxCs & \PSRsGWsRhoMaxCs & \PSRsXrayRhoMaxCs & \AllRhoMaxCs \\
            $p\left(\max c_s^2/c^2\right)$ $[10^{35}\mathrm{dyn}/\mathrm{cm}^2]$ & \PriorPMaxCs & \PSRsPMaxCs & \PSRsGWsPMaxCs & \PSRsXrayPMaxCs & \AllPMaxCs \\
            \hline \hline
        \end{tabular}
    \end{center}
    \caption{
       Marginalized one-dimensional highest-probability credible intervals for selected EoS quantities inferred using current observations.
       We quote the median and 90\% highest-probability-density intervals for the maximum NS mass $\Mmax$, the radius $R_{1.4}$, tidal deformability $\Lambda_{1.4}$, and moment of inertia $I_{1.4}$ of a $1.4\,\Msolar$ NS, along with the the pressure at 1, 2, and 6 times nuclear saturation density.
       We also quote the maximum sound speed attained at any density below the central density of the nonrotating maximum-mass stellar configuration, along with the pressures and densities at which that sound speed is realized.
    }
    \label{tab:ResultsSummary}
\end{table*}

\begin{table*}
    \begin{tabular}{>{\centering}m{0.15\textwidth}>{\centering}m{0.1\textwidth}>{\centering}m{0.1\textwidth}>{\centering}m{0.1\textwidth}>{\centering}m{0.1\textwidth}>{\centering}m{0.1\textwidth}m{0.1\textwidth}}
        \hline \hline
        \multirow{2}{*}{NS} & \multicolumn{2}{c}{$M$ $[\Msolar]$} & \multicolumn{2}{c}{$R$ $[\mathrm{km}]$} & \multicolumn{2}{c}{$\Lambda$} \\
        \cline{2-7}
                                & Initial       & Updated & Initial           & Updated & Initial & Updated \\
        \hline
        GW170817 $m_1$ & \OhEightMone & \UpdatedOhEightMone & \OhEightRone & \UpdatedOhEightRone & \OhEightLone & \UpdatedOhEightLone \\
        GW170817 $m_2$ & \OhEightMtwo & \UpdatedOhEightMtwo & \OhEightRtwo & \UpdatedOhEightRtwo & \OhEightLtwo & \UpdatedOhEightLtwo \\ 
        GW190425 $m_1$ & \OhFourMone & \UpdatedOhFourMone & \OhFourRone & \UpdatedOhFourRone & \OhFourLone & \UpdatedOhFourLone \\
        GW190425 $m_2$ & \OhFourMtwo & \UpdatedOhFourMtwo & \OhFourRtwo & \UpdatedOhFourRtwo & \OhFourLtwo & \UpdatedOhFourLtwo \\
        J0030+0451          & \NicerM & \UpdatedNicerM & \NicerR & \UpdatedNicerR & \NicerL & \UpdatedNicerL \\
        \hline \hline
    \end{tabular}
    \caption{
        The 90\% highest-probability density credible regions for parameters of individual NSs before and after applying our joint constraints on the EoS from the observations of all sources (see Fig.~\ref{fig:MR170817NICER}).
        Initial constraints correspond to the observation of the corresponding system plus the massive pulsar measurements and our EoS prior, while updated constraints include EoS constraints from all observations.
    }
    \label{tab:updated params}
\end{table*}

\section{Projected EoS constraints from future Observations}
\label{sec:EoSfuture}

Discoveries of new sources, or continued observation of existing sources, will enhance the EoS inference reported in the previous section.
In this section, we turn to the projected constraints that could be achievable in the coming $\sim 5$ years. We design and analyze a set of mock observations that mimic what LIGO, Virgo, NICER and other facilities may detect in the near future.

\subsection{Simulated observations}

We assume that the true NS EoS is a specific draw from our EoS prior that is consistent with the current constraints presented in Sec.~\ref{sec:EoScurrent}; it has $R_{1.4}=12.17\,$km, $p(2\rho_{\rm nuc}) = 3.9\times\PtwonucCoef\,\mathrm{dyn}/\mathrm{cm}^2$, $\Lambda_{1.4}=380$, and $\Mmax=2.21\,\Msolar$.
We then use this EoS to simulate upcoming heavy pulsar, GW, and NICER observations. We also consider a future measurement of the MoI of J0737$-$3039A, the primary in the double pulsar system~\cite{BurgayDAmico2003,LyneBurgay2004}.
While the results of our study may depend to a certain degree on the EoS we inject, our choice is a fairly typical example of the EoSs favored by current NS matter knowledge. Our projected constraints could also be enhanced by the observation of electromagnetic counterparts to the BNS coalescences (see e.g.~\cite{Margalit:2017dij}), which we do not consider here.

\begin{table*}
    \begin{tabular}{lccc}
        \hline \hline
       Dataset & Data class & New observations & Total observations \\
        \hline
        \multirow{3}{*}{Mock current}
          & PSR   & 3 real massive pulsars & 3 \\
          & GW    & 1 simulated BNS similar to GW170817 & 1 \\
          & X-ray & 1 simulated $M$-$R$ measurement similar to J0030+0451 & 1 \\
        \hline
        \multirow{3}{*}{End of O4}
          & PSR   & 0 & 3 \\
          & GW    & 4 simulated BNSs distributed as SNR$\sim$SNR$^{-4}$ & 5 \\
          & X-ray & 2 simulated $M$-$R$ measurements, with $M$ measured \textit{a priori} in one case & 3 \\
        \hline
        \multirow{3}{*}{1\,yr at O5}
          & PSR   & 0 & 3 \\
          & GW    & 15 simulated BNSs distributed as SNR$\sim$SNR$^{-4}$ & 20 \\
          & X-ray & 3 simulated $M$-$R$ measurements, with $M$ measured \textit{a priori} in one case & 6 \\
        \hline \hline
    \end{tabular}
    \caption{
        Descriptions of the number and distribution of simulated events in our analysis.
        In addition to the observations listed in the table, we consider the effect of a refined mass measurement for J0740+6620, the discovery of a new $2.20\,\Msolar$ pulsar, and a measurement of the MoI of the double pulsar's primary component.
        See Figs.~\ref{fig:EOScumulmockBNSNICER},~\ref{fig:convergence},~\ref{fig:EOScumulmockMmax}, and ~\ref{fig:EOScumulmockMOI}.
    }
    \label{tab:mock data}
\end{table*}

\subsubsection{Binary neutron star coalescences via gravitational waves}

The ongoing third observing run (O3), and upcoming fourth (O4) and fifth (O5) observing runs, of LIGO and Virgo are expected to yield further GW observations of coalescing NSs~\cite{Prospects2018}. 
In order to simulate such observations, we assume an SNR distribution of SNR$^{-4}$~\cite{2014arXiv1409.0522C} and neglect cosmological effects on this scaling.
Reference~\cite{Lackey:2014fwa} shows that only systems with SNR $\gtrsim 20$ contribute meaningfully to EoS constraints.
Our analysis confirms this, so we restrict our simulation to systems with SNR $> 20$.
Given the expected SNR distribution of sources~\cite{2014arXiv1409.0522C} we find for the number $N$ of systems above a given SNR $N\mathrm{(SNR}>20)/N\mathrm{(SNR}>12) \sim 0.216$.
This fraction can be used to express our projections in terms of the total number of BNS detections and compare them to the expected detection rates for upcoming observing runs.

We assume that NS masses are distributed uniformly between $\Msolar$ and $\Mmax$, while the tidal deformability of each star is computed given the mass and the assumed EoS.
For each simulated signal we approximate the likelihood function for the relevant parameters $({\cal{M}},q,\tilde{\Lambda},\delta\tilde{\Lambda})$ as a Gaussian distribution.
The absolute measurement uncertainty is assumed to be $0.005\ (0.27)$ at the 90\% credible level for ${\cal{M}}\ (q)$ at an SNR of $33$.
For $\tilde{\Lambda}$ we assume a 90\% credible interval of $700$ at SNR $33$, while the likelihood for $\delta \tilde{\Lambda}$ (a different combination of $\Lambda_1$ and $\Lambda_2$ \cite{Favata:2013rwa}) is flat, i.e. the measurement uncertainty is very large~\cite{Wade:2014vqa}.
The above uncertainty values are based on results for GW170817~\cite{Abbott:2018wiz} and scale inversely with the SNR of each signal.
Finally, each likelihood is peaked at the injected value plus a random shift drawn from a Gaussian of width equal to the measurement uncertainty in order to mimic the effect of detector noise.

\subsubsection{X-ray light curves from pulsar hot spots}

Regarding further observations of pulsar hot spots with NICER, we consider the known targets summarized in Table 1 of~\cite{Guillot:2019vqp}, and focus on pulsars for which X-ray oscillations have been detected.
For each pulsar, we use its known mass and uncertainty if available, or else assume a mass of $1.4\,\Msolar$ and a relative measurement uncertainty of $20\%$ at the $68\%$ confidence level.
We then compute the corresponding pulsar radius given our preselected EoS and a measurement error for the pulsar compactness $C$ of $\sigma_C/C=5\%$ inspired by J0030+0451~\cite{Miller:2019cac}. We approximate the likelihood function with a multivariate Gaussian.
Reference~\cite{Raaijmakers:2019qny} also considers the effect of doubling the observation time of J0030+0451 and calculates the ensuing measurement error decrease; it concludes that its effect on EoS constraints will be minimal.\footnote{This assessment may be overly pessimistic because the EoS posterior in \cite{Raaijmakers:2019qny} is already constrained by the narrow choice of prior made in that study. It is possible that continued observation of J0030+0451 could yield further EoS information with respect to a looser prior, such as our \emph{agnostic} one~\cite{Cole_private_comm}.} We therefore do not consider a potential improved constraint from J0030+0451.

\subsubsection{Radio observations of massive pulsars }

Our analysis in Sec.~\ref{sec:EoScurrent} suggests that further measurements of the masses of heavy pulsars around $\sim 2\,\Msolar$ have a minimal effect on EoS constraints.
For this reason we simulate data from only two selected cases.
The first case concerns additional observations of J0740+6620~\cite{Cromartie:2019kug} which could potentially yield a tighter measurement of its mass.
We assume an uncertainty of $0.1\,\Msolar$ at the 68\% confidence level, which could be achieved with ongoing or planned observations~\cite{J0470uncertainty}.
The second simulated case is the potential discovery of a pulsar with a Gaussian-distributed mass of $2.20 \pm 0.044\,\Msolar$ at the 68\% confidence level.
The peak of the posterior for this simulated pulsar is close to the heaviest NS supported by our injected EoS.

\subsubsection{Moment of inertia}

Besides existing EoS probes, we also consider a potential novel measurement, that of the NS MoI via periastron advance of the double pulsar J0737$-$3039. 
Improvements in pulsar timing capabilities may soon allow the periastron advance to be measured with sufficient precision to pick out the correction from spin-orbit coupling, which is proportional to the MoI \cite{KramerWex2009}.
References~\cite{LyneBurgay2004,LattimerSchutz2005,KramerWex2009} predict that this will yield a $\sim 10\%$ measurement of the MoI for the system's primary component.
Here, we make a somewhat more conservative assumption and simulate a Gaussian likelihood for the MoI with $20\%$ precision at the 68\% credible level.
The value of the injected MoI is computed from our assumed EoS based on the known mass of the primary.

\subsection{Results}

\begin{figure}[]
    \parbox{\hsize}{
    \includegraphics[width=\hsize]{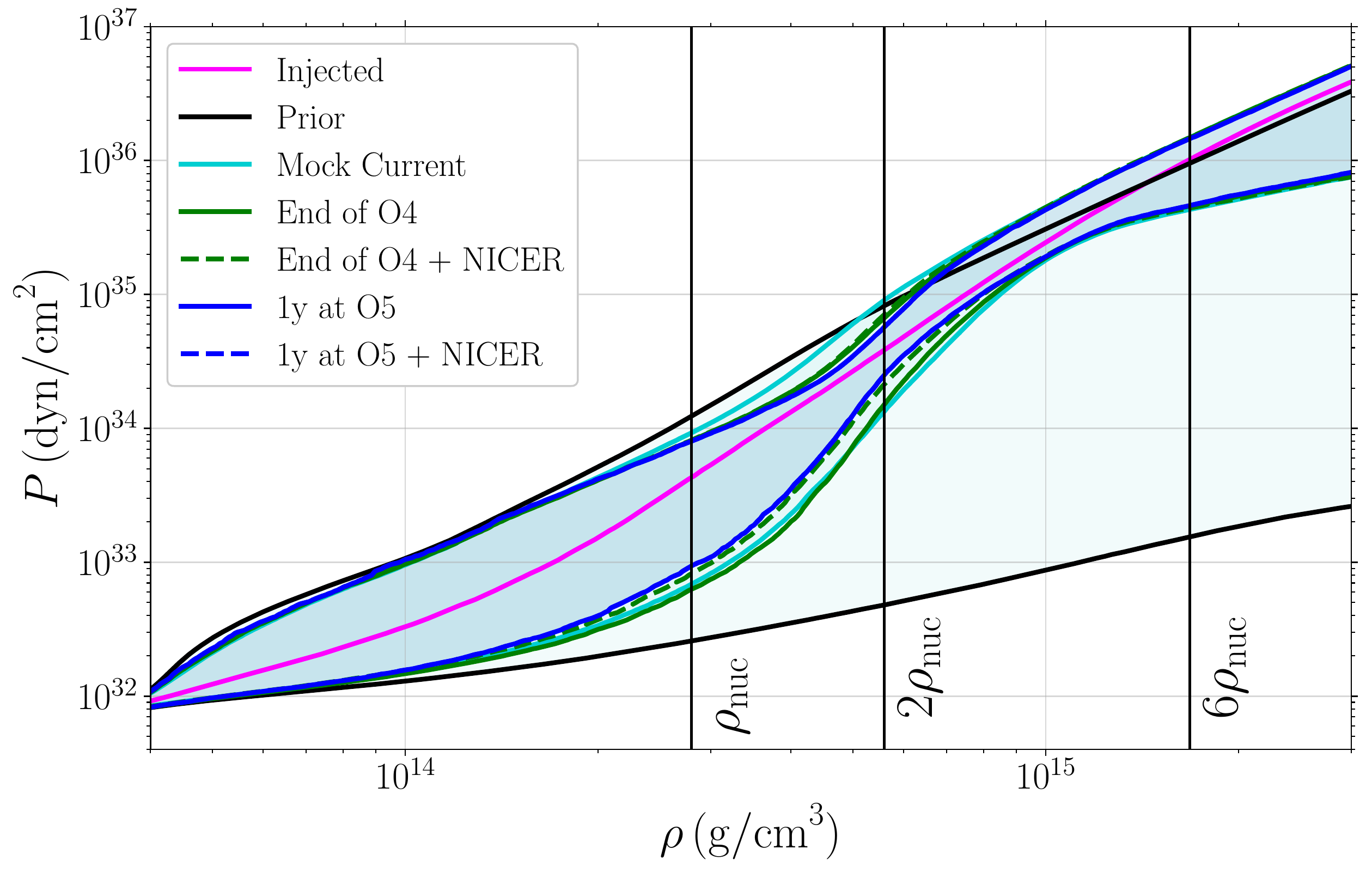}\\
    }
    \parbox{\hsize}{
    \includegraphics[width=\hsize]{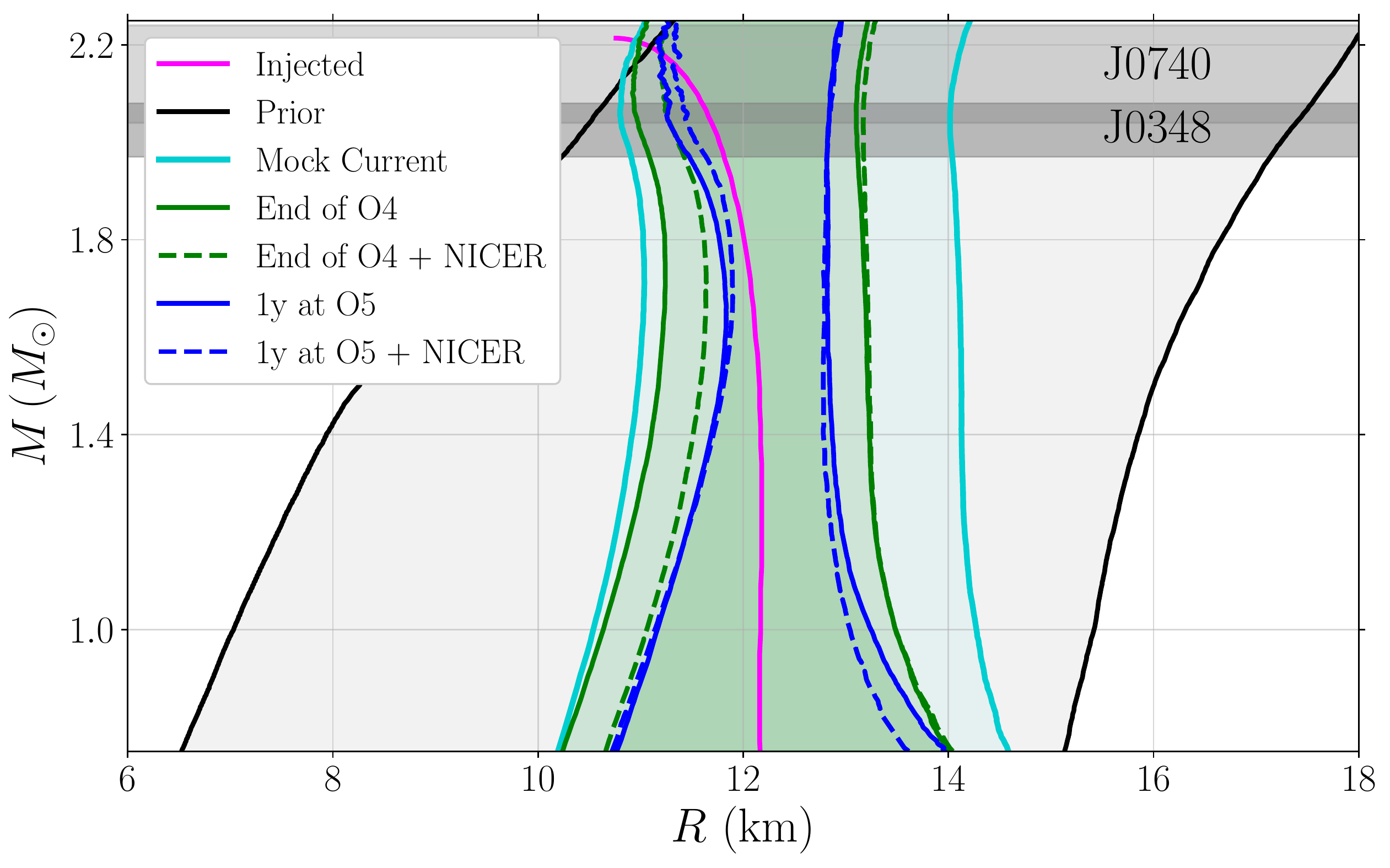}\\[-2ex]
    }
    \caption{
    Projected constraints on the EoS in the pressure-density (top) and mass-radius (bottom) plane from hypothetical future observations. 
    We consider the scheduled LIGO-Virgo observing runs and published NICER targets. 
    Black lines denote our prior range, while the pink line is our injected EoS. The turquoise lines correspond to our mock current constraints given the injected EoS, while
    green and blue solid lines correspond to constraints using potential GWs observed during O4 and O5, respectively. Green and blue dashed lines denote improvements
    over the corresponding solid lines by incorporating potential future NICER results.
    }
    \label{fig:EOScumulmockBNSNICER}
\end{figure}

\begin{figure}[]
    \parbox{\hsize}{
    \includegraphics[width=\hsize]{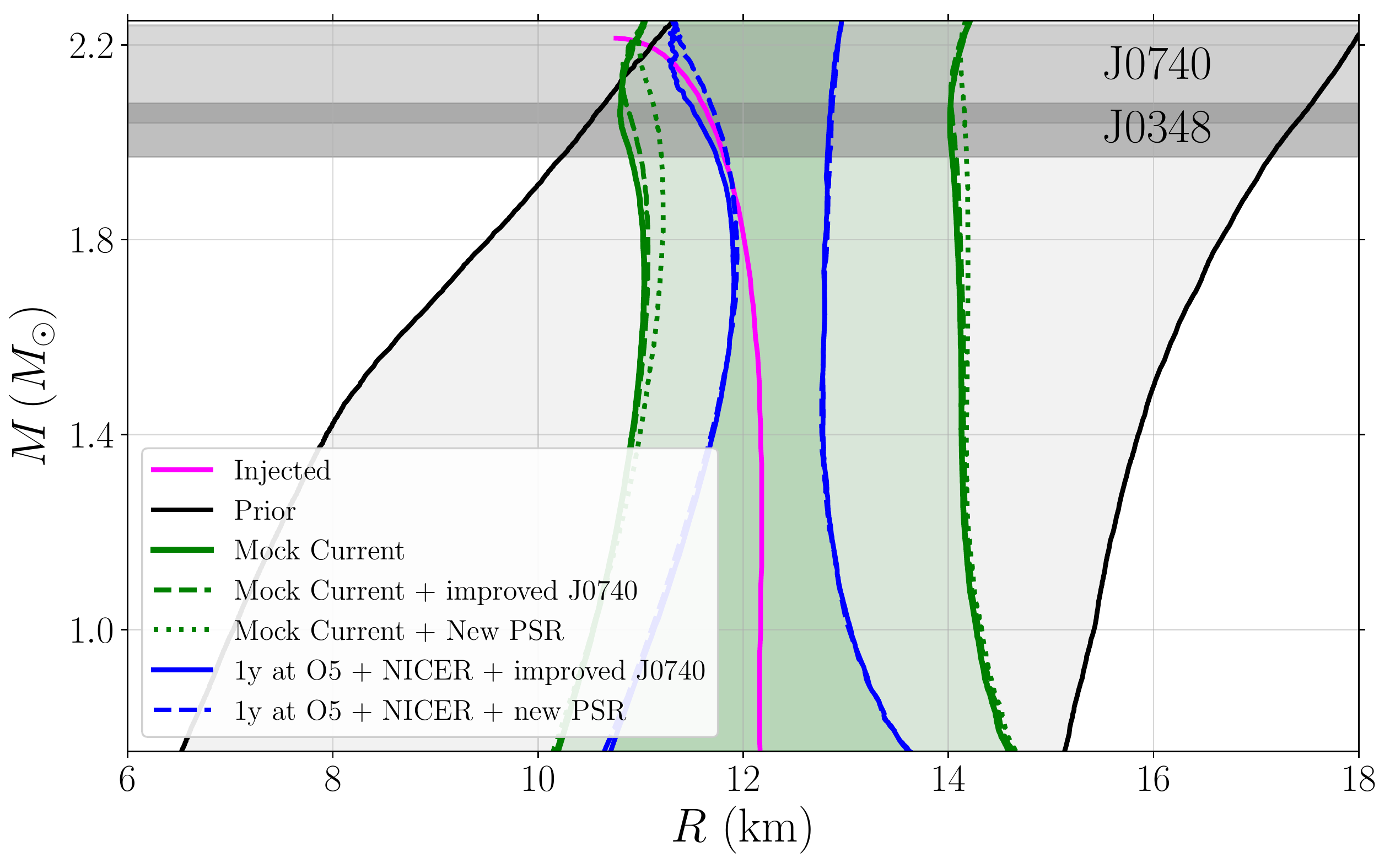}\\[-2ex]
    }
    \caption{
    Impact of further observations of heavy pulsars on the EoS constraints. Black lines denote our prior range, while the pink line is our injected EoS. 
    Green lines correspond to mock current constraints (solid lines) augmented by a potential improved measurement of the mass of J0740+6620 (dashed line) or a potential
    new heavy pulsar observation (dotted line). Blue lines show similar constraints but consider O5-era observations of GWs.
We have omitted the O5-era reference contours that do not include any new pulsar information because they overlap with the solid blue lines.
    }
    \label{fig:EOScumulmockMmax}
\end{figure}

\begin{figure}[]
    \parbox{\hsize}{
    \includegraphics[width=\hsize]{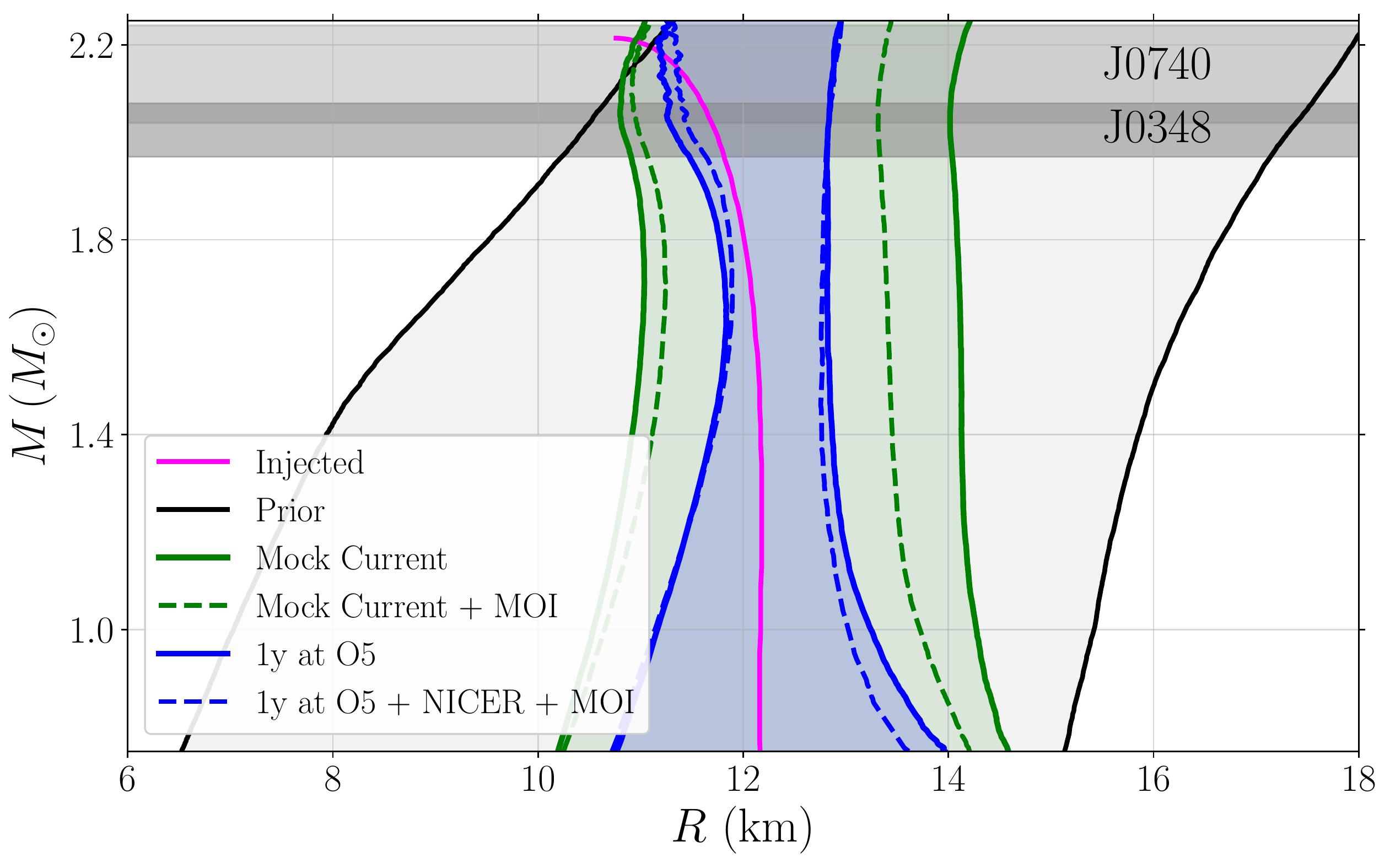}\\[-2ex]
    }
    \caption{
    Impact of a potential future MoI constraint. Black lines denote our prior range, while the pink line is our injected EoS. 
    Green lines correspond to mock current constraints (solid lines) augmented by a potential measurement of the MoI of J0737$-$3039 (dashed lines). 
    Blue lines show similar constraints but consider O5-era observations of GWs and include NICER results.
    }
    \label{fig:EOScumulmockMOI}
\end{figure}

We analyze the simulated data using the same methodology as for the real events.
Figures~\ref{fig:EOScumulmockBNSNICER},~\ref{fig:EOScumulmockMmax}, and~\ref{fig:EOScumulmockMOI} present projected constraints on the EoS from these hypothetical future observations.
Since we cannot be sure whether the EoS we assumed in order to make our simulated data is an accurate representation of the true EoS, for these results we cannot 
employ the current observational constraints computed in Sec.~\ref{sec:EoScurrent} as a starting point.
Instead, we simulate a BNS detection with similar SNR and masses as GW170817, as well as a NICER observation that is comparable to J0030+0451, and use them to compute a mock version of the current constraints (turquoise lines) which we will use as a baseline to compare projected improvements against.
The mock current constraints also include the real observations of massive pulsars, as they are consistent with the injected EoS.

The progressive improvement of EoS constraints as more GWs and NICER observations are added is presented in Fig.~\ref{fig:EOScumulmockBNSNICER}.
We consider the scheduled O4 and O5 LIGO-Virgo observing runs~\cite{Prospects2018}, as well as announced pulsars targeted by NICER~\cite{Guillot:2019vqp}.
The estimated number of BNS detections expected during O4 is $10^{+52}_{-10}$~\cite{Prospects2018}, and we assume that this will result in ${\cal{O}}(4)$ BNS detections with SNR$\,>20$, corresponding to a total number of ${\cal{O}}(10)$ detections.
Projections for O5 are less certain, but given the targeted increases in detector sensitivity, we assume ${\cal{O}}(100)$  BNS detections per year, ${\cal{O}}(10)$ of which have SNR$\,>20$.
For NICER, we combine mock observations of PSR J0751+1807 and PSR J0636+5129 with GW results from O4, and additionally combine PSR J2241$-$5236, PSR J1231$-$1411, and PSR J1012+5307 with GW results from O5.
(The masses for J0751+1807 and J1012+5307 are already known to be $1.64\pm 0.15\,\Msolar$ \cite{LundgrenZepka1995,DesvignesCaballero2016} and $1.83 \pm 0.11 \,\Msolar$ \cite{NicastroLyne1995,AntoniadisTauris2016}, respectively, at the 68\% level.)
The design of our simulated observation campaign is laid out in more detail in Table~\ref{tab:mock data}.

Figure~\ref{fig:EOScumulmockBNSNICER} suggests that the combination of GW and NICER data will result in exquisite EoS constraints in the coming years.
Starting from a mock present-day 90\% credible radius uncertainty $\Delta R_{1.4} \approx \MockCurrentDR\,$km (which is slightly larger than the actual radius uncertainty of $\Delta R_{1.4} \approx \CurrentDR\,$km from Sec.~\ref{sec:EoScurrent}), we find that the discovery of $\sim4$ BNSs in O4 could result in an uncertainty of $\Delta R_{1.4} \approx \MockOFourDR\,$km, for a $\sim\MockOFourDRFraction$ improvement (green solid lines).
Adding information from two pulsars observed by NICER tightens the lower limit on the radius, resulting in $\Delta R_{1.4} \approx \MockOFourNICERDR\,$km, for a $\sim\MockOFourNICERFraction$ improvement over mock current constraints (green dashed lines).
The potential detection of $20$ loud BNSs during O5 can lead to $\Delta R_{1.4} \approx \MockDesignDR\,$km (blue solid lines), while further NICER observations bring the error to $\Delta R_{1.4} \approx \MockDesignNICERDR\,$km (blue dashed lines).
Similarly, from a mock current 90\% credible uncertainty of $\Delta p(2\rho_{\rm nuc}) \approx 7.3\times\PtwonucCoef\,\mathrm{dyn}/\mathrm{cm}^2$ (cf.~the actual $\Delta p(2\rho_{\rm nuc}) = 5.6\times\PtwonucCoef\,\mathrm{dyn}/\mathrm{cm}^2$ from Sec.~\ref{sec:EoScurrent}), the precision of the recovered pressure at twice nuclear saturation density improves by $\sim 34 \%$ after O4 and by a further $\sim 30 \%$ after one year of O5, including the contributions from NICER. 

Not unexpectedly, the radius constraints we obtain are tightest for masses around $1.4$--$1.8\,\Msolar$, as our simulated NICER observations come from this range, while the GW observations are uniformly distributed in mass.
There are relatively few observations of lighter or heavier NSs simply because of the small total number of events we consider. Nonetheless, NSs of $\sim 1\,\Msolar$ may be especially informative as they benefit from stronger tidal interactions. 
The paucity of high-mass NS observations means that we expect to recover only weak constraints on the EoS at densities corresponding to masses above $\sim 2.15\,\Msolar$.

Taking the current observational constraints in Fig.~\ref{fig:EOScumulreal} and the mock results in Fig.~\ref{fig:EOScumulmockBNSNICER} together, we see that GW observations tend to more easily constrain large radii and stiff EoSs, while NICER X-ray observations exhibit the opposite trend.
This is due to the fact that GW observations constrain the tidal interactions in NS binaries, which are more pronounced when the stars have large radii.
They can therefore place stringent upper limits on the radius when tidal effects are not observed.
On the other hand, smaller NSs are more compact and hence result in X-ray light curves that are less variable, suggesting that NICER can more easily place lower limits on the NS radius, even when little variability is resolved.
This complementarity of GW and NICER EoS constraints offers the possibility of ${\cal{O}}(1)$\,km measurements on the NS radius for masses of $\sim 1.4$--$1.8\,\Msolar$ in the coming years, according to our projections.

\begin{figure}[]
    \includegraphics[width=\hsize]{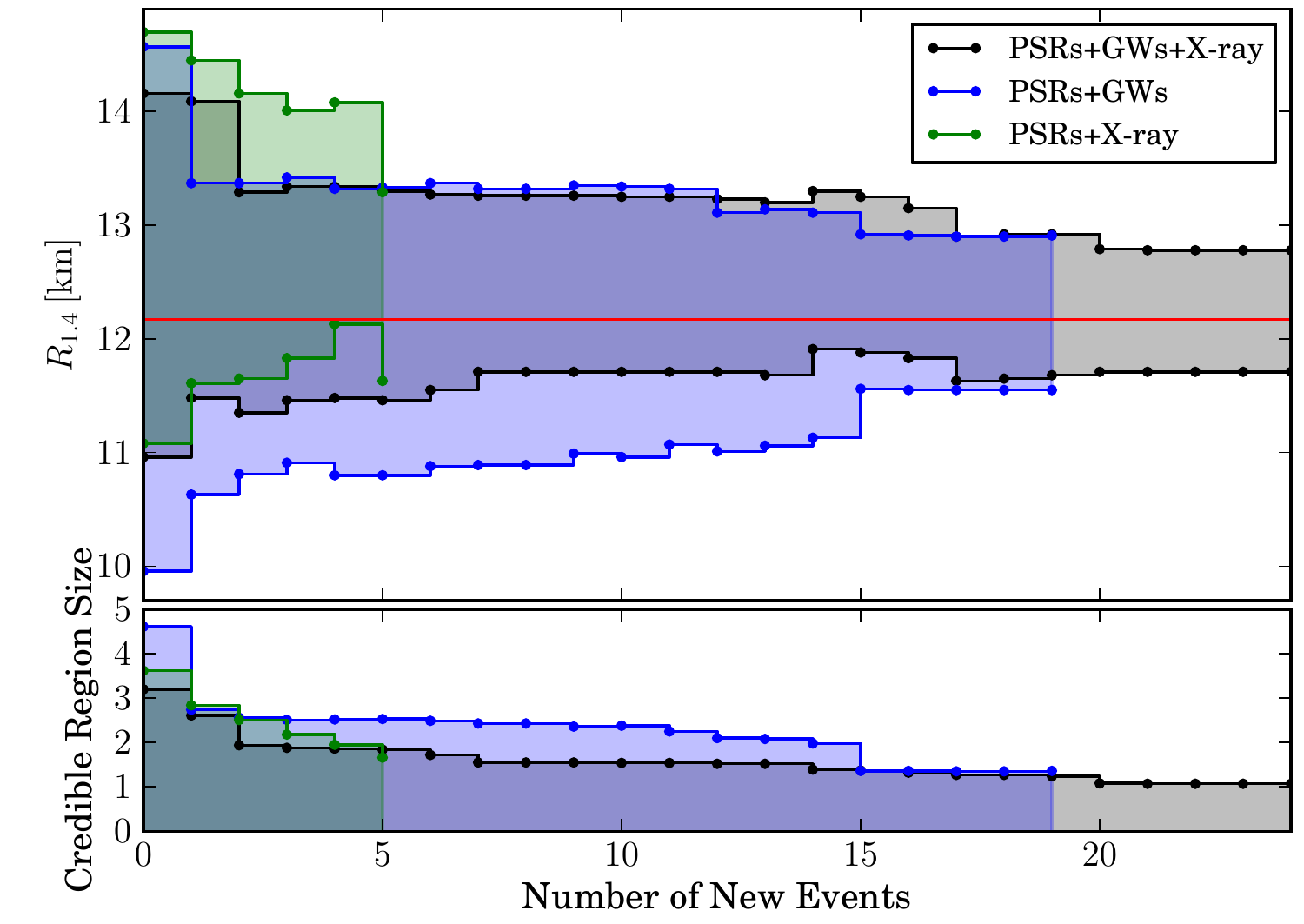}
    \includegraphics[width=\hsize]{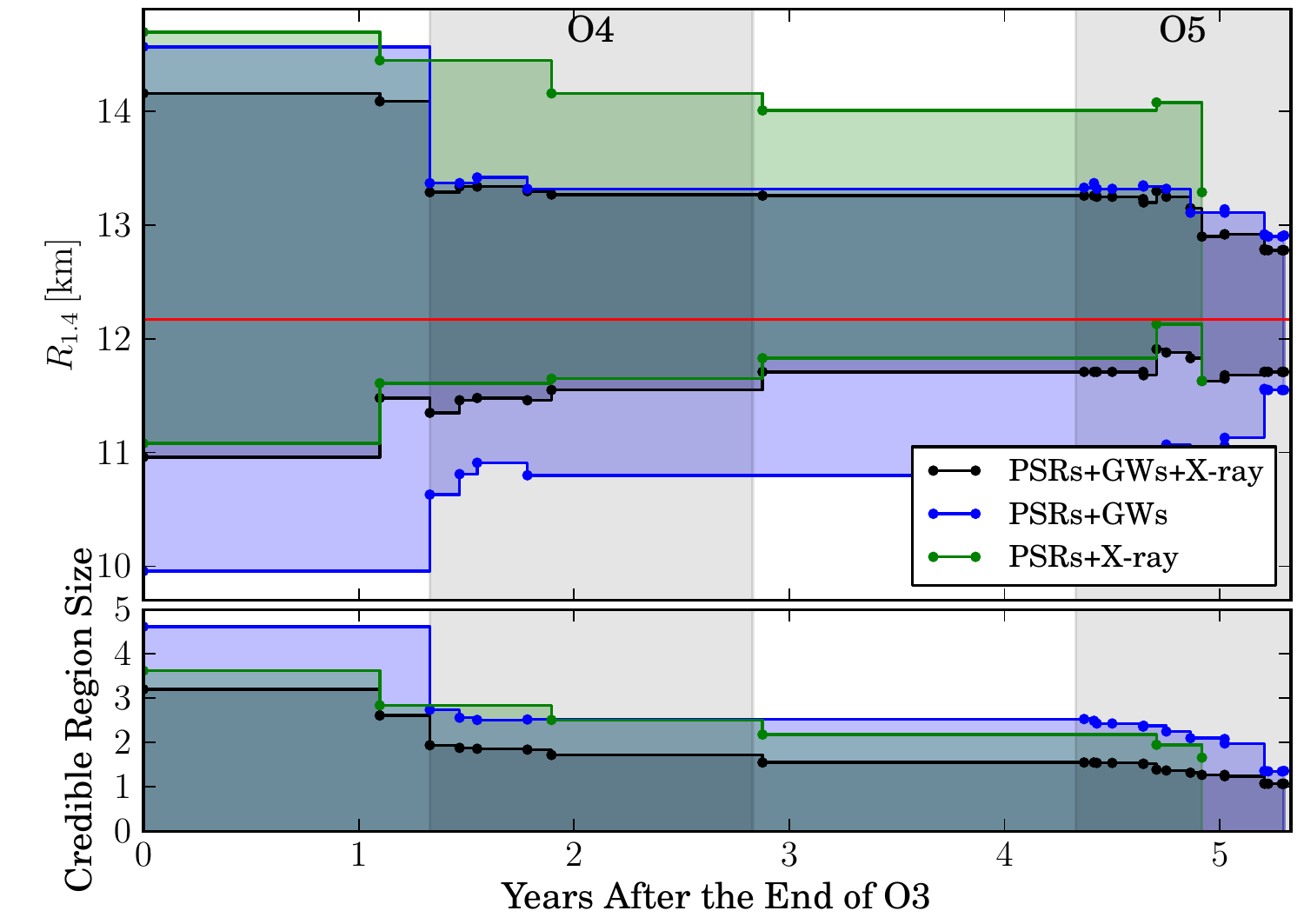}
    \caption{
        Convergence of the 90\% highest-probability-density credible regions for $R_{1.4}$ with the addition of more GW and X-ray observations.
        We find the expected scaling with the number of events (\emph{top}) and show a possible timeline for based on the expected detection rates during O4 and O5 (\emph{bottom}).
        As with current observations, we find that X-ray measurements typically set the lower bound on $R_{1.4}$ while GW observations set the upper bound.
    }
    \label{fig:convergence}
\end{figure}

Although joint constraints that incorporate both new GW and X-ray observations will always be the most stringent, it is also informative to consider the scalings of constraints using only GWs or only X-ray data.
Figure~\ref{fig:convergence} shows this for our simulated observations.
While X-ray observations can produce tighter constraints than GWs for a fixed number of observations, likely due to higher average SNRs for X-ray measurements than for GWs, the fact that more GW detections are expected over the same time period more than compensates for their weaker individual constraints.
GW observations are likely to continue driving the joint constraints for the foreseeable future.
Based on our simulations we expect the size of the 90\% highest-probability-density credible regions for $R_{1.4}$ and $p(2\rho_{\rm nuc})$ to scale with the number of events $N$ as \result{$\Delta R_{1.4} \sim 5.4\,\mathrm{km}/\sqrt{N}$} (\result{$4.0\,\mathrm{km}/\sqrt{N}$}) and \result{$\Delta p(2\rho_{\rm nuc}) \sim 2.8\times\PtwonucCoef\,\mathrm{dyn}\,\mathrm{cm}^{-2}/\sqrt{N}$} (\result{$1.9\times\PtwonucCoef\,\mathrm{dyn}\,\mathrm{cm}^{-2}/\sqrt{N}$}) for GW (X-ray) observations.
The uncertainty in the tidal deformability of a $1.4\,\Msolar$ NS scales as \result{$\Delta\Lambda_{1.4} \sim 910/\sqrt{N}$} (\result{$1200/\sqrt{N}$}).

Besides the expected upcoming observations from GWs and X-rays, other kinds of astronomical observations might also contribute to EoS constraints.
Figure~\ref{fig:EOScumulmockMmax} examines the effect of further heavy pulsar observations on the EoS.
In this plot, green and blue solid lines denote EoS constraints when using only the currently available heavy pulsar measurements from Table~\ref{tab:Maxmassdata}. The corresponding constraints from the additional observations considered here are denoted with dashed and dotted lines.
A better determination of the mass of J0740+6620 or the discovery of a new heavy pulsar might be expected to offer information about the EoS in the high-density regime. 
However, we find negligible overall improvements from additional observations of heavy pulsars, with a reduction in the uncertainty of the radius of a $2.1\,\Msolar$ NS of only \result{$0.2$-$0.3\,$km} compared to today.
Our projected knowledge of $M_{\rm max}$ and $p(6\rho_{\rm nuc})$ in the O5 era remains virtually unchanged relative to the present, whether or not the new observations are included.
In fact, we see that the true EoS lies marginally outside the 90\% credible bound at high masses even after accounting for the new pulsar, due to the fact that it has significant likelihood support ($m=2.20\pm0.044\,\Msolar$) above the true $\Mmax$ of $2.21\,\Msolar$.
Therefore, the new pulsar marginally disfavors the true EoS, particularly compared to the (less constrained) stiffer EoSs that are not ruled out by observations at lower masses.

Finally, we consider a possible measurement of the MoI of J0737$-$3039A in Fig.~\ref{fig:EOScumulmockMOI}.
Green dashed lines show how current EoS constraints could be improved by such a measurement relative to the green solid lines.
We find that $\Delta R_{1.4}\approx \MockCurrentMOIDR\,$km ($\Delta p(2\rho_{\rm nuc}) \approx 5.6\times\PtwonucCoef\,\mathrm{dyn}/\mathrm{cm}^2$), for a $\sim\MockCurrentMOIFraction$ ($\sim 23\%$) improvement compared to current constraints.
However, the MoI measurement has very little impact compared to the O5-era constraints from GWs and X-ray observations.
This is because the MoI measurement acts similarly to an additional GW observation of the tidal deformability~\cite{Yagi:2016bkt}, and the large number of O5-era GW observations tend to overwhelm its contribution.

\section{Discussion}
\label{sec:conclusions}

The EoS inference results presented above update our understanding of NS matter to reflect the information encoded in the latest massive pulsar, GW, and NICER observations.
Relative to previous inferences within the nonparametric framework, we obtain tighter constraints on macroscopic observables such as the canonical radius, which we constrain to be $R_{1.4} = \AllR\,$km, in agreement with the parametric EoS study in \cite{Jiang:2019rcw}.
This improvement of \result{25\%} over the constraint without J0030+0451 reported in \cite{Essick:2019ldf} is possible because of the complementary contributions of GW and NICER observations.

Our study of a mock population of future astronomical observations demonstrates that we might realistically expect to reduce the uncertainty in $R_{1.4}$ by a further \result{50\%} or more over the next 5 years.
Similar gains are expected for the pressure at twice nuclear saturation density and the canonical tidal deformability.
The ultimate $p(2\rho_{\rm nuc})$ constraint we predict is less optimistic than that of \cite{Forbes:2019xaz}, but the relative improvement in $p(2\rho_{\rm nuc})$ knowledge we project is comparable to that predicted by \cite{Lackey:2014fwa} on the basis of $\mathcal{O}(20)$ BNSs following the loudest such signal.
On the other hand, we find that knowledge of the NS maximum mass and the high-density EoS will likely not change significantly on a five-year timescale, in agreement with \cite{Wysocki:2020myz}.
This is primarily because there are few observations of high-mass NSs in our simulated population by virtue of the chosen uniform mass distribution for BNSs coupled with the low total number of events and the fact that the NICER observations are focused on the $1.4$--$1.8\,\Msolar$ range.
Of course, the true population model is unknown, so our conclusion may not be borne out if e.g.~the BNS population and GW selection effects conspire to favor the heaviest NSs.
In any case, it seems likely that a tight constraint on $M_{\rm max}$ will only be achieved in parallel with other population parameters.

Our analysis of the simulated observations also reveals which types of observations will drive the improvements in EoS constraints over the next half-decade.
We find that the BNSs we expect to detect with LIGO and Virgo during O4 and O5 make the biggest contribution to the EoS constraints through sheer numbers.
The simultaneous mass and radius measurements for additional NICER targets are individually very informative, but by the end of O5 they will have little effect on the joint constraints.
Given that, NICER can make the greatest impact by targeting pulsars with masses that differ from those most commonly involved in the BNS coalescences.
In contrast, we find that the discovery of a pulsar more massive than J0740+6620---but still roughly compatible with existing estimates of $M_{\rm max}$---is not particularly helpful for adding to our EoS knowledge.
This is because our nonparametric analysis does not impose strong correlations between the low- and high-density behavior of the EoS.
A NS moment of inertia measurement has the potential to be fairly constraining if incorporated in the near term; by O5, however, it will not make a significant contribution compared to the accumulated GW information.

We find that NICER's preference for stiffer EoSs than GW170817 slightly decreases the already tenuous preference for hybrid stars with multiple stable branches in the $M$-$R$ relation first reported in~\cite{Essick:2019ldf}.
Perhaps because of the greater model freedom allowed by our nonparametric representation of the EoS, we are not able to rule out any phase transition phenomenology based on the joint astronomical observations, although the connections between the constraints on macroscopic EoS observables and the nuclear microphysics merit further investigation (see e.g.~\cite{Ferreira:2019bgy,ZimmermanCarson2020}). Conversely, nuclear theory predictions, e.g.~from chiral effective field theory~\cite{HebelerSchwenk2010,TewsKruger2013,LynnTews2016,DrischlerCarbone2016}, are known to be helpful in constraining the low-density EoS~\cite{Tews2018b,JiangTang2019,EssickTews2020}. Indeed, \cite{Christian:2019qer} studied the question of hybrid stars for a specific class of parametrized phase transitions and found that NICER and GW170817 together are inconsistent with a strong phase transition at low densities when information from chiral effective field theory is taken into account, whereas \cite{EssickTews2020} did not find compelling evidence that phase transitions are forbidden when using chiral effective field theory with our nonparametric EoS representation.

Finally, through our metholodogy in Sec.~\ref{sec:methods}, we have laid out a blueprint for progressively incorporating new observational information into a nonparametric inference of the EoS as it becomes available. While we focused on a five-year horizon in this study, the same methods can be applied to make projections over the longer term. The massive pulsar, GW, and X-ray observations studied here will undoubtedly continue to shape our understanding of NS matter in the coming years.

\section*{Acknowledgments}

The authors thank Cole Miller, Jocelyn Read, Andrew Steiner, and Sukanta Bose for helpful suggestions about this work.
P.~L. is supported by National Science Foundation Grant No.~PHY-1836734 and by a gift from the Dan Black Family Trust to the Gravitational-Wave Physics \& Astronomy Center.
R.~E. is supported at the University of Chicago by the Kavli Institute for Cosmological Physics through an endowment from the Kavli Foundation and its founder Fred Kavli.
The Flatiron Institute is supported by the Simons Foundation. 
The authors are grateful for computational resources provided by the LIGO Laboratory and supported by National Science Foundation Grants No.~PHY-0757058 and No.~PHY-0823459.
This analysis was made possible by the \textsc{numpy}~\cite{numpy} and \textsc{matplotlib}~\cite{Hunter:2007} software packages.

\appendix

\section{Mass priors that depend on the EoS}
\label{sec:Occam factors}

As mentioned at several points in Sec.~\ref{section:eos inference}, the mass priors for each type of dataset depend on both the underlying population and the EoS, 
taking the form $P(m|\varepsilon, \lambda) $.
In particular, if the observed compact object is known to be a NS, then its mass $m$ cannot exceed the maximum mass supported by the EoS, $\Mmax$.
Generally, this implies a prior of the form
\begin{equation}
    P(m|\varepsilon, \lambda) = \frac{P(m|\lambda) \Theta(m\leq\Mmax)}{\int\limits^{\Mmax} dm'\, P(m'|\lambda)},
\end{equation}
where we explicitly renormalize the prior that is determined by the population model $P(m|\lambda)$, and $\lambda$ denotes the relevant population parameters.
If the astrophysical population model for compact objects predicts masses that are strictly below $\Mmax$ for this EoS, i.e.~if there is no astrophysical formation scenario that can create NSs as heavy as possibly allowed by the EoS, then there is no net effect on the prior and $P(m|\varepsilon, \lambda) = P(m|\lambda)$.
However, if the population model supports compact objects with masses larger than $\Mmax$, then there is a nontrivial normalization that depends on the EoS. 
This is a type of Occam factor as it depends on the prior volume of each EoS, the amount of parameter space over which it has support \textit{a priori}.
Failing to incorporate the appropriate prior normalization term would result in an incorrect population model, and it is known that assuming an incorrect population model can bias the inferred EoS.

To further elucidate this point, we consider a toy model in which our analysis assumes that the population model predicts a flat distribution between $M_\mathrm{min}$ and up to 
$M_\mathrm{pop} > \Mmax$.
In this case, the mass prior becomes
\begin{equation}
    P(m|\varepsilon, \lambda) = \left\{ \begin{matrix} \frac{1}{\Mmax - M_\mathrm{min}} & \text{ iff } & M_\mathrm{min} \leq m \leq \Mmax, \\ 0 & \text{ else, } & \end{matrix} \right.
\end{equation}
where $\Mmax$ depends on $\varepsilon$.
Furthermore, we assume that the mass measurement is perfect so that $P(d|m) = \delta(m-m_\mathrm{true})$, where $m_\mathrm{true}$ is the mass of the observed NS.
If the true population model does not match the population model we assumed in our analysis, but instead it only produces objects up to $M^\mathrm{true}_\mathrm{pop} < M_\mathrm{pop}$, then any EoS with $\Mmax > M^\mathrm{true}_\mathrm{pop}$ will have a likelihood
\begin{align}
    P(d|\varepsilon) & = \int dm \, P(m|\varepsilon, \lambda) \delta(m - m_\mathrm{true}) \nn \\
                     & = (\Mmax(\varepsilon) - M_\mathrm{min})^{-1}<1.
\end{align}
The likelihood of this EoS is decreased by an amount that depends on the maximum mass $\Mmax$ it predicts.
As a result, an EoS with $\Mmax$ only slightly larger than $m_\mathrm{true}$ is favored over an EoS with larger $\Mmax$, even though they both explain the observed data equally well.

If we perform a joint analysis with many events and the incorrect population model, we will strongly favor $\Mmax\sim M^\mathrm{true}_\mathrm{pop}$, even though the real EoS may have a maximum mass $\Mmax^\mathrm{true} \gg M^\mathrm{true}_\mathrm{pop}$.
This bias in the inferred EoS is directly attributable to our faulty assumption about the population.
If the assumed population model is correct, then any EoS with $\Mmax \ge M_\mathrm{pop}=M^\mathrm{true}_\mathrm{pop}$ will be considered equally likely, and no bias will be introduced, reemphasizing the need to consider both the mass distribution and the EoS simultaneously when analyzing many observations.

\section{Additional results with real events}
\label{sec:moreplots}

\begin{figure*}[]
    \parbox{0.48\hsize}{
    \includegraphics[width=\hsize]{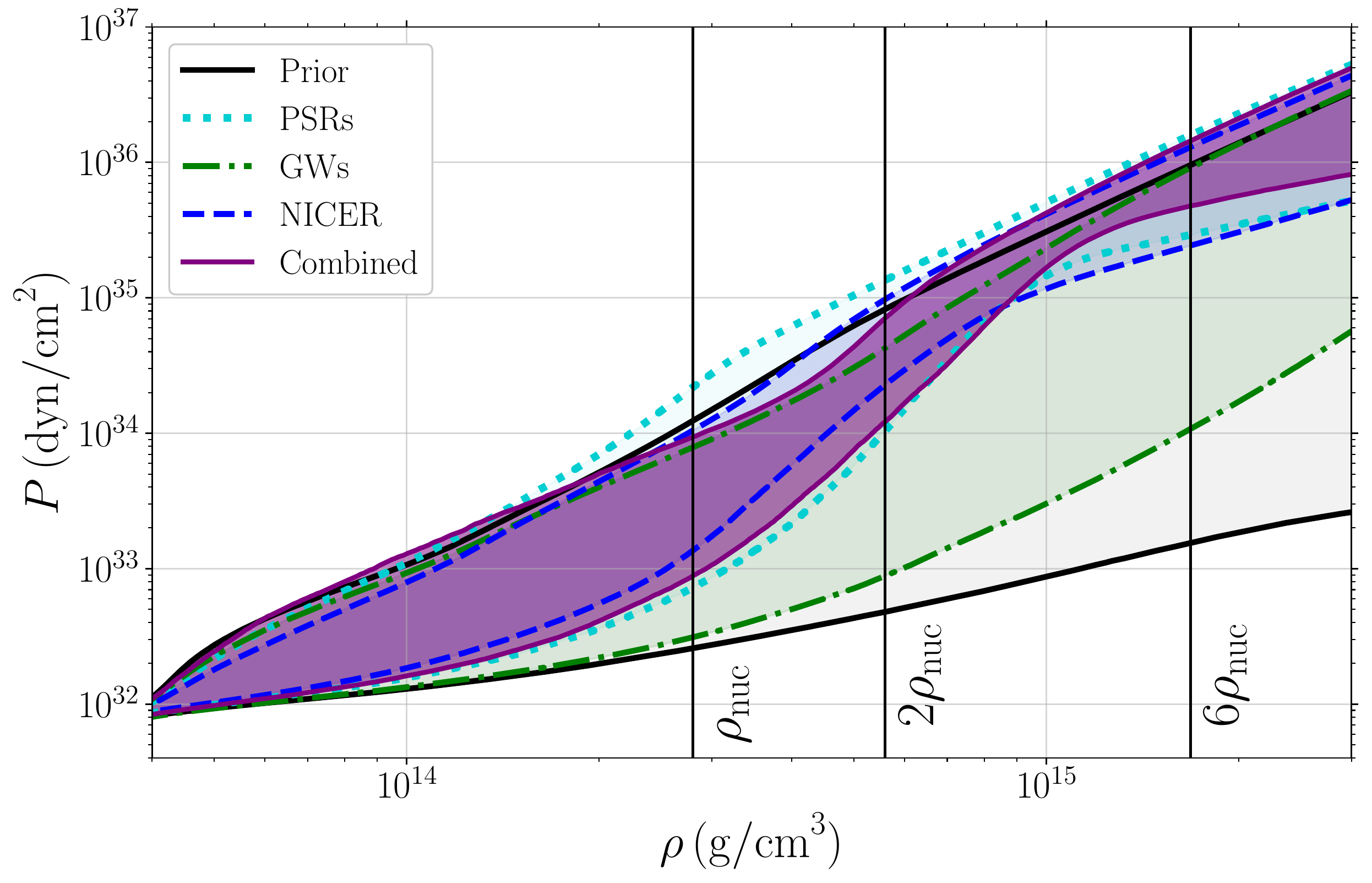}\\[-2ex]
    }\parbox{0.48\hsize}{
    \includegraphics[width=\hsize]{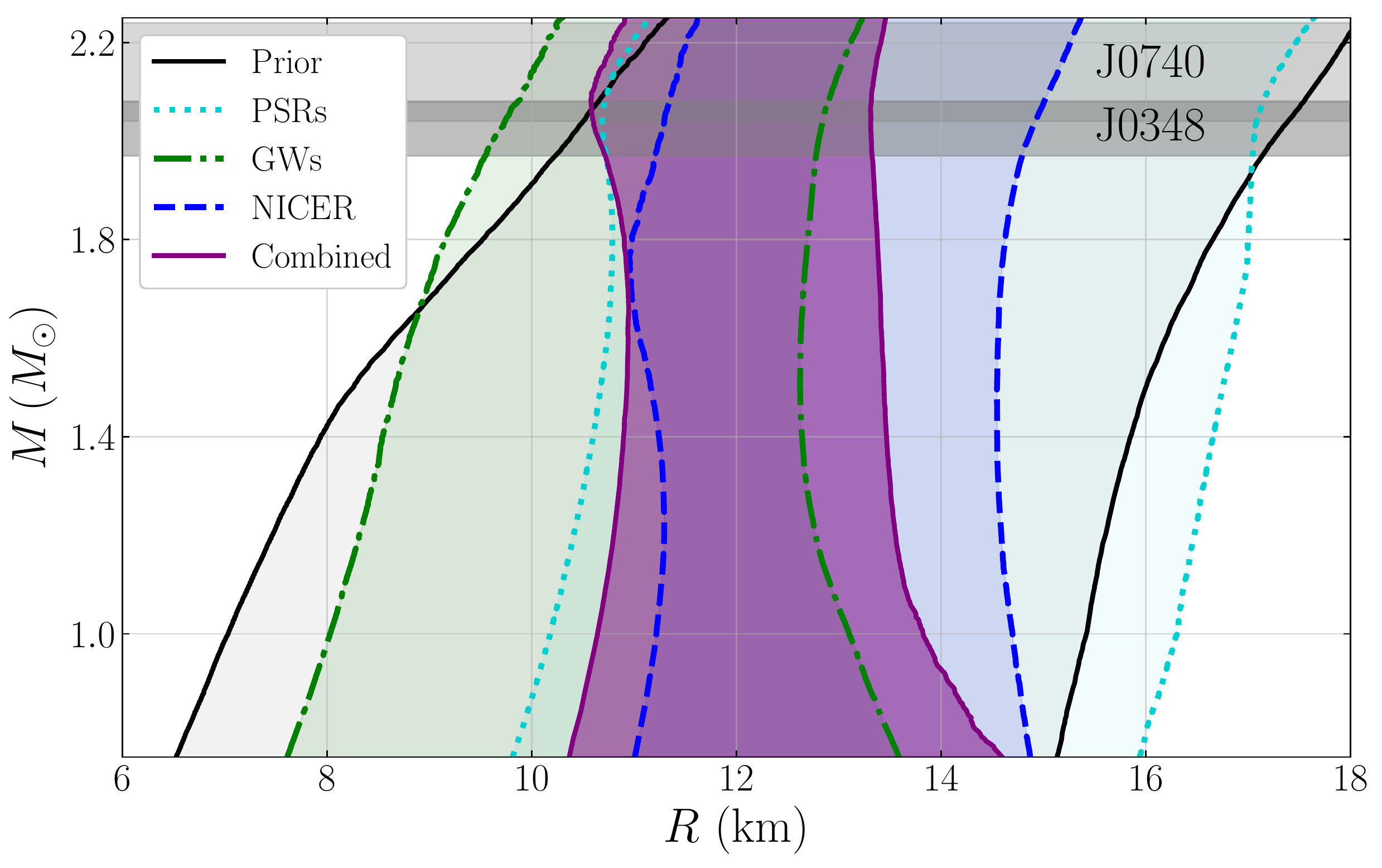}\\[-2ex]
    }
    \caption{
        Similar to Fig.~\ref{fig:EOScumulreal}, but with individual-event rather than cumulative constraints on the EoS from different classes of observations.
        We show 90\% credible intervals for the EoS in the pressure-density (left panel) and mass-radius (right panel) planes. 
        Black lines denote the prior range while turquoise lines correspond to the posterior when using only the heavy pulsar measurements.
        The shaded regions correspond to the posterior when only employing the GW data (green) or NICER (blue).
        Vertical lines in the left panel denote multiples of the nuclear saturation density, while horizontal shaded regions in the right panel show the 68\% credible mass estimate for the two heaviest known pulsars.
      }
    \label{fig:EOSindivreal}
\end{figure*}

In this appendix we present additional results for the analysis that constrains the EoS using current observational data.
Figure~\ref{fig:EOSindivreal} shows 90\% symmetric credible intervals in the pressure-density (left panel) and mass-radius (right panel) planes from individual observational constraints, as opposed to the cumulative constraints of Fig.~\ref{fig:EOScumulreal}.
These plots again confirm that GW observations alone (green region) cannot presently be used to place a strong lower limit on the NS radius, but instead result in an upper limit of $\sim 13\,$km.
Conversely, the analysis of J0030+0451 results in a lower limit of $\sim 11\,$km for the radius, showcasing the nice complementarity between GW and X-ray observations of NSs.
The NICER measurement appears to be more informative in our inference than in \cite{Miller:2019nzo,Raaijmakers:2019dks,Raaijmakers:2019qny} because those works adopt a much narrower EoS prior, both in extent and in allowed phenomenological behavior. References~\cite{Raaijmakers:2019dks,Raaijmakers:2019qny} additionally count the massive pulsar data as part of the prior. The 90\% credible region of their $M$-$R$ prior falls almost entirely within the corresponding region of our J0030+0451 posterior in Fig.~\ref{fig:EOSindivreal}.

\begin{figure*}[]
    \includegraphics[width=0.9\textwidth]{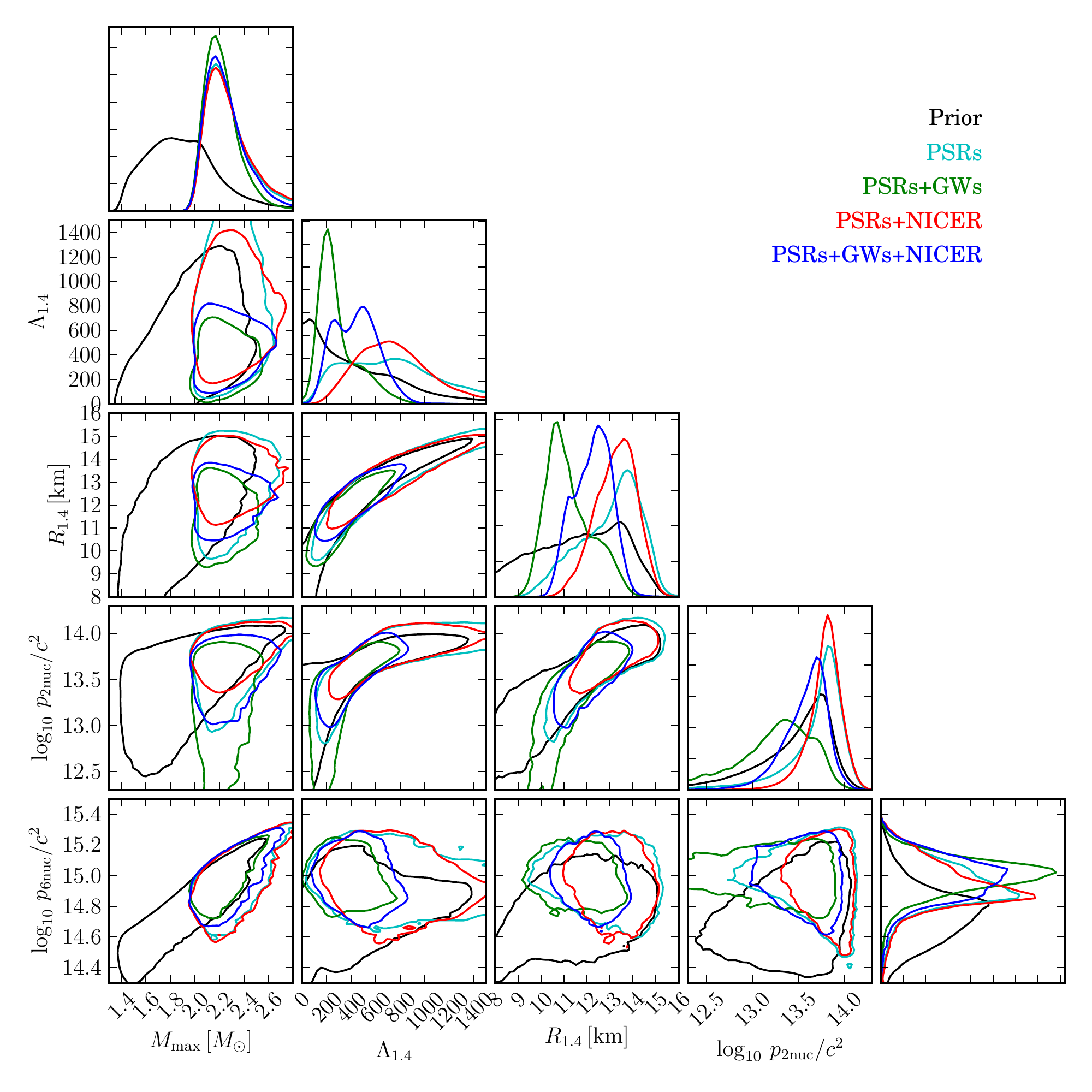}
    \caption{
        Correlations between $\Mmax$, $\Lambda_{1.4}$, $R_{1.4}$, and the pressure at twice and six times nuclear saturation (in units of dyn/cm$^2$).
        Contours in the joint distributions correspond to 90\% credible intervals.
        We show results from our \emph{agnostic} prior (\emph{black}), the observation of massive pulsars (\emph{cyan}, Table~\ref{tab:Maxmassdata}), massive pulsars and GW observations (\emph{green}, Tables~\ref{tab:Maxmassdata} and~\ref{tab:BNSdata}), massive pulsars and NICER observations (\emph{red}, Tables~\ref{tab:Maxmassdata} and~\ref{tab:NICERdata}), and all observations together (\emph{blue}, Tables~\ref{tab:Maxmassdata},~\ref{tab:BNSdata}, and~\ref{tab:NICERdata}), matching the credible intervals reported in Table~\ref{tab:ResultsSummary}.
    }
    \label{fig:corner}
\end{figure*}

Figure~\ref{fig:corner} shows correlations between selected microscopic and macroscopic NS properties for analyses using different combinations of observational constraints.
We find that the posterior for the maximum mass is primarily driven by the heavy pulsars, and GW or NICER observations offer little additional 
information.
Additionally, we confirm the known correlation between $\Mmax$ and the pressure at six times nuclear saturation density, present both in our prior and all our posteriors.
At the same time, $\Mmax$ is not strongly correlated with the pressure at lower densities.

Inference about these lower densities is primarily driven by the lower-mass GW and NICER observations.
As expected, we find that $R_{1.4}$ and $\Lambda_{1.4}$ are correlated with the pressure at twice the nuclear saturation density.
Examining the one-dimensional posteriors, we again conclude that GW observations point toward smaller NS radii and tidal parameters, while NICER observations have the opposite effect.
The joint constraint comes from the union of both measurements.

\section{Comparison to the Riley \textit{et al.} analysis}
\label{sec:Riley}

\begin{figure}[]
    \parbox{\hsize}{
    \includegraphics[width=\hsize]{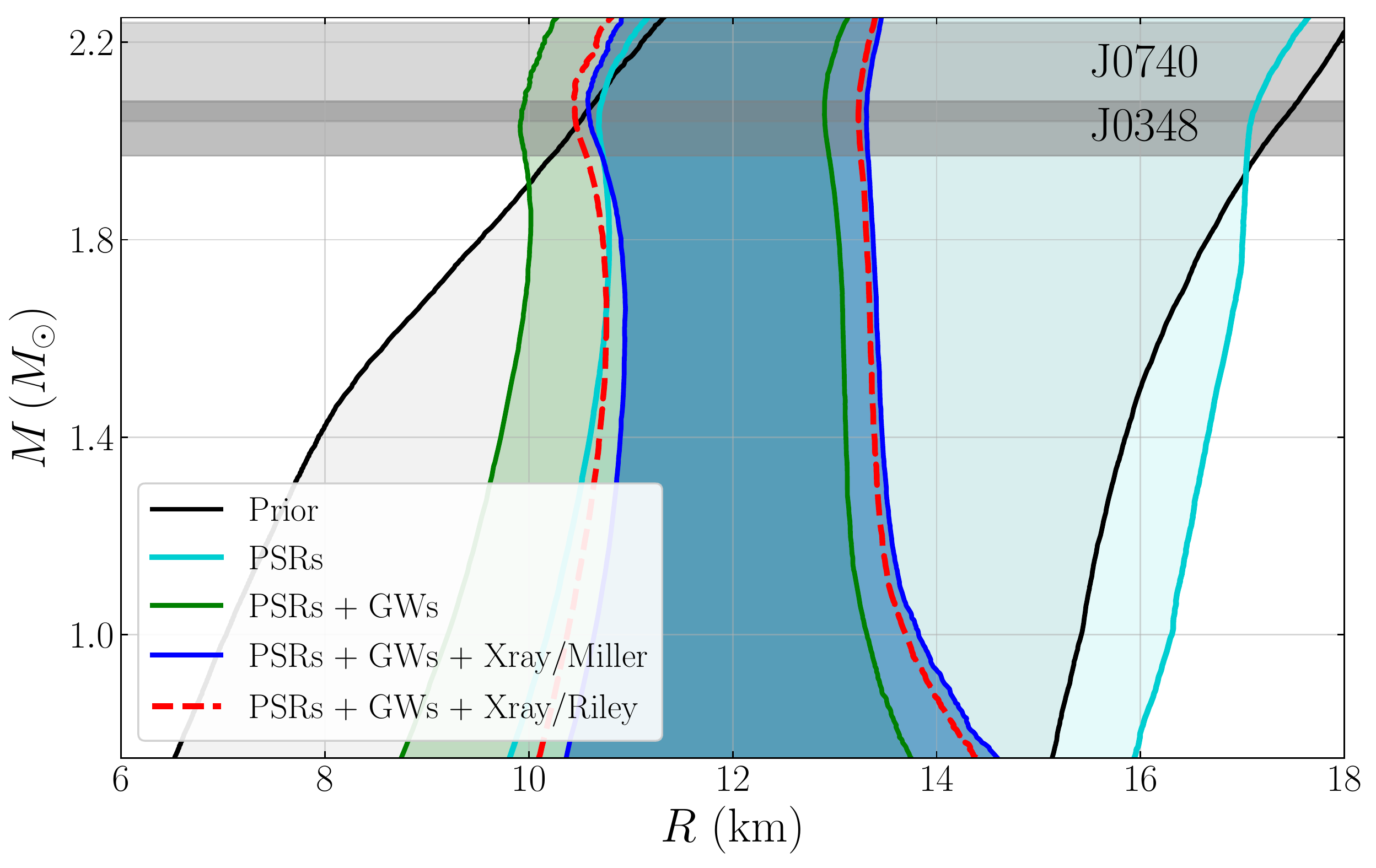}\\[-2ex]
    }
    \caption{
    Similar to the right panel of Fig.~\ref{fig:EOScumulreal}, with the addition of the red dashed lines that denote the combined 90\% credible interval in the mass-radius plane when using the analysis of Riley \textit{et al.}~\cite{Riley:2019yda}. We find that the Miller \textit{et al.}~\cite{Miller:2019cac} and the Riley \textit{et al.}~\cite{Riley:2019yda} analyses are consistent within their statistical errors.
    }
    \label{fig:EOScumulrealRiley}
\end{figure}

The main results presented in this work make use of the Miller \textit{et al.}~\cite{Miller:2019cac} analysis of the NICER data.
In this appendix, we show that the independent analysis of Riley \textit{et al.}~\cite{Riley:2019yda} leads to a consistent inference of the properties of the EoS.
Figure~\ref{fig:EOScumulrealRiley} shows cumulative constraints on the mass-radius plane similar to the right panel of Fig.~\ref{fig:EOScumulreal}.
In Fig.~\ref{fig:EOScumulrealRiley} we also display the combined 90\% credible interval when instead using results from the ST+PST model of Riley \textit{et al.}~\cite{Riley:2019yda} (see Table~\ref{tab:NICERdata}).
Inferred values for selected EoS parameters are presented in Table~\ref{tab:AltResultsSummary}.
Overall, the two independent analyses of Miller \textit{et al.}~\cite{Miller:2019cac} and Riley \textit{et al.}~\cite{Riley:2019yda} are consistent within their statistical errors.

\begin{table*}[]
    \begin{center}
        \begin{tabular}{l @{\quad}r @{\quad}r}
            \hline \hline
            Observable              & PSRs+X-ray/Riley    & PSRs+GWs+X-ray/Riley \\
            \hline
            $\Mmax$ $[\Msolar]$    & \AltPSRsXrayMmax      & \AltAllMmax \\
            $R_{1.4}$ $[{\rm km}]$ & \AltPSRsXrayR         & \AltAllR \\
            $\Lambda_{1.4}$       & \AltPSRsXrayL         & \AltAllL \\
            $I_{1.4}$ $[10^{45}\,\mathrm{g}\,\mathrm{cm}^2]$
                                  & \AltPSRsXrayI         & \AltAllI \\
            $p(\rho_\mathrm{nuc})$ $[\PnucCoef{\rm dyn}/{\rm cm^2}]$       & \AltPSRsXrayPnuc      & \AltAllPnuc \\
            $p(2\rho_\mathrm{nuc})$ $[\PtwonucCoef{\rm dyn}/{\rm cm^2}]$   & \AltPSRsXrayPtwonuc   & \AltAllPtwonuc \\
            $p(6\rho_\mathrm{nuc})$ $[\PsixnucCoef{\rm dyn}/{\rm cm^2}]$   & \AltPSRsXrayPsixnuc   & \AltAllPsixnuc \\
            $\max c_s^2/c^2$ & \AltPSRsXrayMaxCs & \AltAllMaxCs \\
            $\rho\left(\max c_s^2/c^2\right)$ $[10^{15}\mathrm{g}/\mathrm{cm}^3]$ & \AltPSRsXrayRhoMaxCs & \AltAllRhoMaxCs \\
            $p\left(\max c_s^2/c^2\right)$ $[10^{35}\mathrm{dyn}/\mathrm{cm}^2]$ & \AltPSRsXrayPMaxCs & \AltAllPMaxCs \\
            \hline \hline
        \end{tabular}
    \end{center}
    \caption{
       Marginalized one-dimensional credible intervals for selected EoS quantities inferred using Riley \textit{et al.}~\cite{Riley:2019yda} instead of Miller \textit{et al.}~\cite{Miller:2019cac} for NICER observations.
       We quote the median and 90\% highest-probability-density intervals for the maximum NS mass $\Mmax$, the radius $R_{1.4}$, tidal deformability $\Lambda_{1.4}$, and moment of inertia $I_{1.4}$ of a $1.4\,\Msolar$ NS, along with the pressure at 1, 2, and 6 times nuclear saturation density.
       We also quote the maximum sound speed attained at any density below the central density of the nonrotating maximum-mass stellar configuration, along with the pressures and densities at which that sound speed is realized.
       Compare to Table~\ref{tab:ResultsSummary}.
    }
    \label{tab:AltResultsSummary}
\end{table*} 

\bibliography{refs}

\end{document}